\documentclass[fleqn,usenatbib,nofootinbib]{mnras}
\usepackage{graphicx}
\usepackage{amsmath,amssymb}
\usepackage{lastpage}
\usepackage[all]{hypcap}
\usepackage[T1]{fontenc}
\usepackage{float}
\usepackage{subfigure}
\usepackage{afterpage}
\usepackage[usenames]{color}
\usepackage{mathrsfs}
\usepackage{upgreek}
\usepackage{hyperref}
\usepackage{gensymb}
\usepackage{color}
\usepackage[dvipsnames]{xcolor}
\usepackage{longtable}
\usepackage{threeparttable}
\usepackage{multirow}
\usepackage{placeins}
\usepackage{bm}
\usepackage[capitalise]{cleveref}
\usepackage{enumitem}

\newcommand{\go}{g_\text{obs}}
\newcommand{\gb}{g_\text{bar}}
\newcommand{\Ug}{\Upsilon_\text{gas}}
\newcommand{\Ud}{\Upsilon_\text{disk}}
\newcommand{\Ub}{\Upsilon_\text{bulge}}

\newcommand{\sig}{\sigma_\text{int}}
\newcommand{\e}{e_\text{N}}

\title[The underlying RAR]{The underlying radial acceleration relation}
\author[H.~Desmond]{
Harry~Desmond$^{1}$\thanks{E-mail: harry.desmond@port.ac.uk}
\\
$^{1}$Institute of Cosmology \& Gravitation, University of Portsmouth, Dennis Sciama Building, Portsmouth, PO1 3FX, UK\\\\
}

\pubyear{2023}

\begin{document}
\label{FirstPage}
\pagerange{\pageref{FirstPage}--\pageref{LastPage}}
\maketitle

\begin{abstract}
The radial acceleration relation (RAR) of late-type galaxies relates their dynamical acceleration, $\go$, to that sourced by baryons alone, $\gb$, across their rotation curves.
Literature fits to the RAR have fixed the galaxy parameters on which the relation depends---distance, inclination, luminosity and mass-to-light ratios---to their maximum a priori values with an uncorrelated Gaussian contribution to the uncertainties on $\gb$ and $\go$. In reality these are free parameters of the fit, contributing systematic rather than statistical error.
Assuming a range of possible functional forms for the relation with or without intrinsic scatter (motivated by Modified Newtonian Dynamics with or without the external field effect),
I use Hamiltonian Monte Carlo to perform the full joint inference of RAR and galaxy parameters for the Spitzer Photometry and Accurate Rotation Curves (SPARC) dataset. This reveals the intrinsic RAR underlying that observed.
I find an acceleration scale $a_0=(1.19 \pm 0.04 \, \text{(stat)} \pm 0.09 \, \text{(sys)}) \: \times \: 10^{-10}$ m s$^{-2}$, an intrinsic scatter $\sig=(0.034 \pm 0.001 \, \text{(stat)} \pm 0.001 \, \text{(sys)})$ dex (assuming the SPARC error model is reliable) and weak evidence for the external field effect. I make summary statistics of all my analyses publicly available for future SPARC studies or applications of a calibrated RAR, for example direct distance measurement.
\end{abstract}


\section{Introduction}
\label{sec:intro}


Galaxies are observed to follow several tight and regular scaling relations between their internal motions and morphology. The classical correlations are the Tully--Fisher relation between rotation velocity and mass or luminosity in late-type galaxies (e.g.~\citealt{Tully-Fisher,McGaugh_BTFR,Pizagno}) and the Fundamental Plane relating luminosity, size and velocity dispersion (e.g.~\citealt{FP_1,FP_2,Cappellari}, including its projection onto the mass--velocity plane, the Faber--Jackson relation;~\citealt{Faber-Jackson}) in early types. These are largely subsumed in late-type galaxies by the mass discrepancy--acceleration or radial acceleration relation~\citep{Milgrom_1, MDAR_Sanders, MDAR_McGaugh, RAR}, relating the local total acceleration, $\go$, to that sourced by baryons, $\gb$, across rotation curves. This provides more detailed radial information about the gravitational potential.

These relations provide the key evidence concerning the \emph{mass discrepancy problem} in galaxies, namely that the motions of stars and gas imply far higher dynamical than baryonic masses in a Newtonian analysis. In the prevailing $\Lambda$ Cold Dark Matter ($\Lambda$CDM) cosmology this difference is assumed to be made up by dark matter, leading to attempts to explain the relations through the modelling of galaxy formation, the galaxy--halo connection and halo mass distributions (e.g.~\citealt{Gnedin,Blanton,Desmond_TFR,Desmond_FJFP, DC_Lelli, Ludlow, Navarro, Keller, Desmond_MDAR, Tenneti, Paranjape_Sheth}).
However, the fact that galaxy formation in $\Lambda$CDM proceeds in a highly stochastic and complicated manner may make it difficult to explain ``simple'' (power-law or roughly double power-law) dynamical scaling relations.

An alternative hypothesis is that Newtonian gravity breaks down at the galaxy scale. Surprisingly, galaxy dynamics can be explained well by a model in which $\go=\gb$ for $\gb \gg a_0$ and $\go\propto\gb^{1/2}$ as $\gb \ll a_0$, where $a_0 \approx 10^{-10}$ m s$^{-2}$ is a new fundamental constant. This naturally leads to the observed simplicity in the aforementioned scaling relations. Supplemented by an ``interpolating function'' that connects the Newtonian and modified gravity regimes, this theory is known as Modified Newtonian Dynamics (MOND; \citealt{Milgrom_1, Milgrom_2, Milgrom_3}) and has achieved some success at explaining and even predicting galaxy behaviour (e.g.~\citealt{Famaey_McGaugh, McGaugh_Milgrom, Paper_I}). MOND has been incorporated into a range of nonrelativistic and relativistic theories over the past four decades, as reviewed most recently in \citet{Banik}.

The RAR is MOND written in terms of observables for late-type galaxies. This at once gives the relation central importance in the missing mass debate and makes it the most sensitive probe of gravitational parameters within the MOND paradigm. These include the acceleration constant $a_0$ marking the onset of modified dynamics, the intrinsic scatter $\sig$ and possibly a parameter $\e \equiv g_\text{ext}/a_0$ describing the influence of mass surrounding the galaxy (external field effect, EFE; \citealt{Milgrom_1}), where $g_\text{ext}$ is the strength of the gravitational field in which the galaxy is embedded. $\sig$ bears on the question of whether the RAR manifests law-like gravitational behaviour as posited by MOND, and also determines the precision with which the relation may be used to calibrate galaxy properties such as distance (analogously to the Tully--Fisher relation). $\e$ addresses the key question within the MOND paradigm of the extent to which---and manner in which---modified gravity or inertia violates the strong equivalence principle. Previous fits have found $a_0 \approx 1.2\times10^{-10}$ m/s$^2$, $\e \approx 0.003$ and a small intrinsic scatter $\sig<0.1$ dex \citep{RAR,Li,Paper_II,Paper_III}. The existence of the EFE is however by no means well-established (for example \citealt{Hernandez} and \citealt{Freundlich} find evidence against it), and qualitatively similar phenomenology may arise in $\Lambda$CDM~\citep{PS_EFE}.

$\gb$ and $\go$ depend on a number of properties of the galaxies, most importantly their distance $D$, inclination $i$, luminosity $L$ and mass-to-light ratios $\Upsilon$ of their various components. These are nuisance parameters when determining the properties of the RAR, although of course of interest in their own right. Past RAR studies have either fixed these to their maximum a priori values given other measurements and then propagated their uncertainties into $\gb$ and $\go$ as if they were random and uncorrelated \citep{RAR}, or varied both the nuisance \emph{and} RAR parameters galaxy-by-galaxy, effectively assuming a different RAR for each galaxy \citep{Li,Paper_I,Paper_II,Paper_III}. Assuming an underlying universal form for the RAR, a superior inference constrains \emph{global} RAR parameters along with the local galaxy properties. The main advantage of this is that it propagates the prior distributions of the galaxy parameters as systematic rather than statistical uncertainties, thus capturing the correlations across rotation curves that fluctuations in these parameters induce. For example, a higher (lower) $\Upsilon$ than expected in a particular galaxy causes a higher (lower) $\gb$ across its rotation curve, yet modelling it as a statistical uncertainty implicitly assumes that a fluctuation in $\Upsilon$ could scatter $\gb$ up at one point and down at the next. The full inference also captures the degeneracies between the RAR and galaxy parameters, which have a non-trivial impact on the relation through the shape of the galaxy priors. This is the analysis I perform here.

Although conceptually simple, the full inference is technically challenging because it implies a vastly higher-dimensional parameter space than the simplified versions. The analysis of \citet{RAR} has two parameters ($a_0$, $\sig$), while that of \citet{Li} has four ($D$, $i$, $\Ud$, $\Ub$) repeated $N=147$ times for $N$ galaxies. (\citealt{Paper_I} additionally sample $\Ug$ and $\e$.) \citeauthor{Li} and~\citeauthor{Paper_I} cannot accommodate parameters that couple the galaxies, so fix $a_0=1.2\times10^{-10}$ m s$^{-2}$ a priori and can at best reconstruct $\sig$ post-hoc from the distribution of residuals, thus neglecting its degeneracy with the other variables. The full inference has up to $6N+n+2$ parameters ($a_0$, $\sig$, $N\times\e$, $N\times D$, $N\times i$, $N\times L_{3.6}$, $N\times \Ud$, $N\times\Ug$, $n\times\Ub$) where $n=31$ is the number of galaxies with bulges.
Thus, although the total number of parameters that I sample is only slightly larger than \citeauthor{Paper_I}, the fact that I sample them \emph{together} while \citeauthor{Li} and~\citeauthor{Paper_I} split them by galaxy makes for a qualitatively different analysis, capable of mapping out the degeneracies between all parameters and inferring $a_0$ and $\sig$.
915 parameters is indeed beyond many sampling methods, but routine for Hamiltonian Monte Carlo. This will enable a robust determination of the RAR parameters for arbitrary priors and assumptions about the underlying functional form. $\gb$ and $\go$ transformed according to the best-fit galaxy parameter values (Fig.~\ref{fig:transformed}) reveals the RAR that underlies the sampling distributions of those parameters. 

The structure of this paper is as follows. In Sec.~\ref{sec:data} I describe the Spitzer Photometry and Accurate Rotation Curves (SPARC) data and selection criteria I employ. Sec.~\ref{sec:method} gives the methodology, including the likelihood model, priors, treatment of the galaxy parameters and details of the sampler. The results are presented in Sec.~\ref{sec:results}. Sec.~\ref{sec:disc} discusses the broader ramifications of the study, remaining systematic uncertainties and useful further work, while Sec.~\ref{sec:conc} concludes. Throughout, $\log$ has base 10 and accelerations are given in $10^{-10}$ m s$^{-2}$ unless otherwise stated.

\section{Observational Data}
\label{sec:data}


I analyse the SPARC sample \citep{SPARC},\footnote{\url{http://astroweb.cwru.edu/SPARC/}} comprising 175 rotation curves from the literature with \textit{Spitzer} photometry at 3.6$\mu$m. I apply the quality cuts recommended by \citet{RAR}, removing galaxies with quality flag 3 (indicating large asymmetries, non-circular motions and/or offsets between stellar and H\textsc{i} distributions) or maximum a priori $i<30\deg$, and points for which the quoted fractional uncertainty on the observed rotation velocity is greater than 10 per cent. This leaves $2696$ points from $147$ galaxies, of which all have mass in a stellar disk but only $31$ have mass in a central bulge.

Distances are determined by a variety of methods with a corresponding range of uncertainties \citep{SPARC}, while the inclinations are estimated from tilted-ring fits to the velocity fields. I use these as Gaussian priors in the inference. The total luminosity at 3.6$\:\mu$m, $L_{3.6}$, is well-measured but its uncertainty is quoted so I include it as a Gaussian prior for completeness and to eliminate statistical uncertainty in the independent ($\gb$) direction which complicates the inference (see Sec.~\ref{sec:inference}). I follow the SPARC convention that $L_{3.6}$ is calculated using the maximum a priori distance for each galaxy, $\bar{D}$, and hence does not scale with $D$. Similarly, the uncertainty on $L_{3.6}$, $\delta L_{3.6}$, comes purely from the uncertainty on the flux and does not include a contribution from the distance uncertainty. The disk and bulge mass-to-light ratios, $\Ud$ and $\Ub$, are believed to be $\sim0.5$ and $\sim0.7$ respectively, with a $\sim25$ per cent uncertainty~\citep{Meidt,McGaugh-Schombert,SPARC}. I use these as lognormal priors, which are marginally favoured over Gaussian given the way the parameters are determined (S.~McGaugh and F.~Lelli, priv. comm.). $\delta L_{3.6}$ is sufficiently small for it not to make a difference whether it is modelled as normal or lognormal.

With H\textsc{i} mass measured, a correction factor must be applied to calculate the total gas mass and hence the gas contribution to $\gb$. The fiducial SPARC analysis uses a conversion factor of 1.33 (accounting for primordial helium), but a more accurate determination includes a scaling of the hydrogen fraction with the stellar mass $M_*$ of the galaxy~\citep{research_note}:
\begin{equation}
M_\text{gas} = X^{-1} \: M_\text{HI}
\end{equation}
where
\begin{equation}
X(M_*) = 0.75 - 38.2 \: \left(M_* / (1.5\times10^{24} M_\odot)\right)^{0.22}.
\end{equation}
As the H\textsc{i} mass has already been scaled by 1.33 in SPARC, I define
\begin{equation}
\bar{\Upsilon}_\text{gas}(M_*) = 1/(1.33 \: X(M_*))
\end{equation}
where overbar denotes maximum a priori value. ($M_*$ must be determined after sampling $L_{3.6}$, $\Ud$ and $\Ub$.) This scales $\Ug$ relative to the value assumed in SPARC when calculating $V_\text{gas}$, as $\Ud$ and $\Ub$ do for $V_\text{disk}$ and $V_\text{bul}$. The results are not significantly altered compared to $\bar{\Upsilon}_\text{gas}=1$. $\Ug$ is given a lognormal prior with 10 per cent width \citep{SPARC}.\footnote{To convert between normal and lognormal distributions I use the full equations relating their means and standard deviation
\begin{eqnarray}\label{eq:log_transform}
&&\mu = \exp(\tilde{\mu}+\tilde{\sigma}^2/2),\\
&&\sigma^2 = (\exp(\tilde{\sigma}^2)-1) \exp(2 \tilde{\mu} + \tilde{\sigma}^2),\nonumber
\end{eqnarray}
with inverse
\begin{eqnarray}\label{eq:log_transform_inv}
&&\tilde{\mu} = \ln(\mu) - \ln(1+\sigma^2/\mu^2)/2,\\
&&\tilde{\sigma}^2 = \ln(1+\sigma^2/\mu^2),\nonumber
\end{eqnarray}
where a tilde indicates the lognormal. The uncertainties are sufficiently small in most cases for this not to differ appreciably from the more common first-order approximation.}

\section{Method}
\label{sec:method}

\begin{figure}
  \centering
  \includegraphics[width=0.5\textwidth]{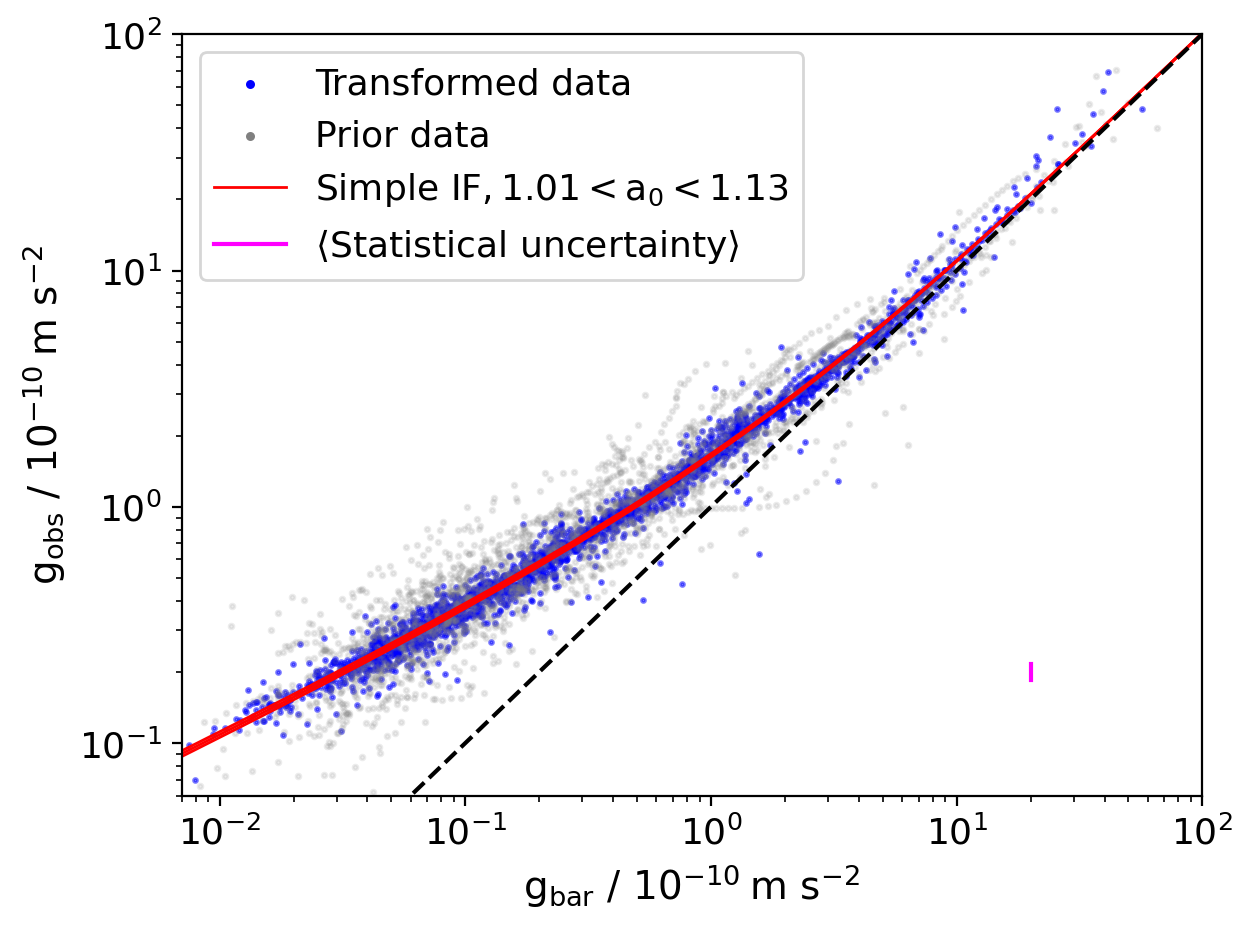}
  \caption{The underlying RAR of the SPARC sample (blue) is obtained by transforming $\gb$ and $\go$ according to the best-fit galaxy parameters, in this case those at the median of the posterior for the inference with intrinsic scatter but without EFE. The Simple IF fit with $a_0$ in its $2\sigma$ allowed range is overplotted in red, and the median errorbar size, deriving solely from the statistical uncertainty in $V_\text{obs}$, is shown as a magenta bar in the lower right. The standard ``prior RAR'', where the galaxy parameters take their maximum a priori values, is shown in faded grey.}
  \label{fig:transformed}
\end{figure}


\subsection{Modelling the RAR}
\label{sec:model}

I fit two functions to the RAR. The first is the ``Simple interpolating function (IF)'' \citep{simple}:
\begin{equation}\label{eq:simple_if}
\go^\text{pred} = \gb/2 + \sqrt{\gb^2/4 + \gb \: a_0}.
\end{equation}
Although in tension with Solar System measurements this function is highly successful for galaxy dynamics \citep{Famaey_McGaugh}, and may readily be tweaked to circumvent local constraints without appreciably altering its larger-scale behaviour. One such modification is the ``RAR IF'' of \citet{RAR}, which I have checked yields almost identical results to the Simple IF. The IF currently has no physical significance and must be constrained empirically \citep{Milgrom_2016, Famaey_McGaugh}.

The reason I use the Simple IF is that the second function I consider is designed to reduce to it in the zero-external-field limit. This is the EFE formula for the nonrelativistic AQUAdratic Lagrangian (AQUAL;~\citealt{AQUAL}) theory of MOND designed in \citet{Chae_Milgrom}:

\scriptsize
\begin{flalign}\label{eq:efe}
&\go^\text{pred} = \gb \left(\frac{1}{2} + \left(\frac{1}{4} + \left(\left(\frac{\gb}{a_0}\right)^2 + (1.1 e_\text{N})^2\right)^{-\frac{1}{2}}\right)^\frac{1}{2}\right) \times \nonumber \\
&\Biggl(1 + \tanh\left(\frac{1.1 e_\text{N}}{\gb/a_0}\right)^{1.2} \times \left(-\frac{1}{3}\right) \times \\
&\frac{\left(\left(\left(\frac{\gb}{a_0}\right)^2 + (1.1 e_\text{N})^2\right)^{-\frac{1}{2}}\right) \left(\frac{1}{4} + \left(\left(\frac{\gb}{a_0}\right)^2 + (1.1 e_\text{N})^2\right)^{-\frac{1}{2}}\right)^{-\frac{1}{2}}}{1 + \left(\frac{1}{2} + 2\left(\left(\frac{\gb}{a_0}\right)^2 + (1.1 e_\text{N})^2\right)^{-\frac{1}{2}}\right)^\frac{1}{2}}\Biggr), \nonumber
\end{flalign}
\normalsize
where $\e \equiv g_\text{ext}/a_0$ describes the strength of the external field at the galaxy in question. The EFE arises in most formulations of MOND due to the theory's nonlinearity: the strong equivalence principle is violated because the acceleration of a system as a whole cannot be transformed away in calculation of its internal motions. This implies that otherwise identical galaxies in different gravitational environments have different kinematics. A stronger external field pushes the system towards the Newtonian regime by reducing the gravitational boost of MOND,
causing a downturn in the RAR at low $\gb$ where $g_\text{ext}$ can be a non-negligible fraction of $\go$. While several fitting formulae for the EFE exist (e.g. \citealt{Banik_EFE, Haghi, EFE_QUMOND}), Eq.~\ref{eq:efe} is the most sophisticated in allowing for variable disk thickness and scale length---and the orientation of the field relative to the disk axis through azimuthal averaging---and has been shown to yield good agreement with the SPARC data \citep{Paper_III, Chae_distinguishing}. It should be borne in mind however that this does not make it correct in general.

I consider both the case of $\e$ as a global parameter describing the average external field over the sample, and as a parameter varying galaxy-by-galaxy to describe their separate local environments. In the former case I use a uniform prior sufficiently broad to enclose the full posterior; in the latter, where there is insufficient information in the data for a meaningful constraint on $\e$, I impose a prior based on the environmental field estimates of the SPARC galaxies from~\citet{Desmond_reconstructing, Paper_II}. These are determined entirely independently of the SPARC data by summing contributions to the gravitational field from the baryonic masses of surrounding objects, including a sophisticated treatment of survey incompleteness and other missing mass. 

As my fiducial analysis I use the results assuming that missing baryons are strongly clustered around visible objects (``maximum clustering'') because this is expected in MOND and was shown in \citet{Paper_II,Paper_III} to give good agreement with the SPARC rotation curves. I also consider an ``average clustering'' model that assumes a prior distribution midway between the ``max clustering'' and ``no clustering'' (missing baryons uncorrelated with visible objects) results, with a width given by half the difference between the two. This systematic uncertainty is larger than the statistical uncertainty in either clustering case separately. The most precise calculation of \citet{Paper_II} uses data from the Sloan Digital Sky Survey and hence is only valid within the footprint of that survey, which includes 90 galaxies in my sample. For the remaining 57 I take $\bar{e}_\text{N}$ to be the median $\e$ over all SPARC galaxies (in the corresponding clustering model), with an uncertainty twice the median uncertainty for all SPARC galaxies. This corresponds to a conservatively wide prior for galaxies without object-specific prior information, while still leveraging information on the $\e$ distribution across the population. Combined with the no-EFE case (Eq.~\ref{eq:simple_if}), these EFE models ought roughly to span the space of possible EFE behaviour and hence indicate the level of systematic uncertainty that the unknown EFE behaviour induces.


Table~\ref{tab:params} summarises the free parameters of the inference and their priors.

\subsection{Inference procedure}
\label{sec:inference}

The parameters inferred in the fiducial model are $a_0, \sig, 147\times\e, 147\times D, 147\times i, 147\times L_{3.6}, 147\times \Ud, 147\times \Ug$ and $31\times \Ub$. At any point in parameter space I calculate $\gb$ and $\go$ as
\begin{equation}\label{eq:gbar_scaling}
\gb = \frac{\left(\Ug V_\text{gas}\:|V_\text{gas}| + L_{3.6}/\bar{L}_{3.6} \: (\Ud V_\text{disk}^2 + \Ub V_\text{bul}^2)\right)}{r},
\end{equation}
\begin{equation}\label{eq:gobs_scaling}
\go = \frac{V_\text{obs}^2}{r} \: \frac{\sin(\bar{i})^2}{\sin(i)^2} \: \frac{\bar{D}}{D},
\end{equation}
where $V_\text{gas}$, $V_\text{disk}$, and $V_\text{bul}$ are the velocities generated by the gas, disk and bulge, $V_\text{obs}$ is the observed velocity and $r$ is the galactocentric radius. These are as quoted in the SPARC database, i.e. assuming $D=\bar{D}$, $i=\bar{i}$, $L_{3.6}=\bar{L}_{3.6}$ and all $\Upsilon=1$. $V_\text{gas}\:|V_\text{gas}|$ is used rather than $V_\text{gas}^2$ in $\gb$ to account for the possibility of central ``holes'' in the gas distribution which can cause the gravitational field sourced by the gas to point outwards. Note that $\gb$ is independent of $D$ because all of $V_\text{gas}^2$, $V_\text{disk}^2$, $V_\text{bul}^2$ and $r$ scale proportionally to $D$.

The only remaining uncertainty to treat as statistical is the contribution of $\delta V_\text{obs}$ to $\go$.\footnote{This is a combination of a formal error from the entire disk fit and a contribution from the difference between the velocities of the approaching and receding sides of the disk (see \citealt{SPARC}, eq. 1). These noise terms are not uncorrelated Gaussian random variables, so a further improvement to the method would be to either sample them or model their covariance structure (see Sec.~\ref{sec:disc}).} I assume this is lognormal, so that
\begin{equation}
\delta \log(\go) = \sqrt{\log(1+(\delta \go/\go)^2)}/\ln(10)
\end{equation}
where
\begin{equation}
\delta \go/\go = 2\:\delta V_\text{obs}/V_\text{obs}.  
\end{equation}
I then use either Eq.~\ref{eq:simple_if} or~\ref{eq:efe} to calculate the predicted $\go$ at each $\gb$. $\sig$ simply adds in quadrature with $\delta \log(\go)$, so the likelihood is:
\begin{flalign}\label{eq:likelihood}
\mathcal{L}(d|\vec{p}) = & \prod_j \frac{1}{\sqrt{2\pi\sigma_\text{tot}^2}} \times \\
& \exp\{-(\log(g_\text{obs,j})-\log(g_\text{obs,j}^\text{pred}))^2/(2\:\sigma_\text{tot,j}^2)\} \nonumber
\end{flalign}
where $\vec{p}$ is the parameter vector and
\begin{equation}\label{eq:scatter_1}
\sigma_\text{tot,j}^2 \equiv \delta\log(g_\text{obs,j})^2+\sig^2.
\end{equation}
$j$ runs over the $2696$ data points.

This fiducial analysis assumes no statistical uncertainty on the velocities sourced by the gas, disk and bulge. However the calculation of $V_\text{disk}$, $V_\text{bul}$ and $V_\text{gas}$ in \citet{SPARC} made assumptions about the 3D geometry of these baryons, particularly in the thickness of the disk components. Variation may be expected to alter the baryon velocities at the $\sim$10-15 per cent level (F. Lelli, priv. comm.). I therefore also consider models in which these velocities are each assigned 10 per cent uncorrelated Gaussian uncertainties,\footnote{These assumptions are unlikely to hold in detail because, as with the other nuisance parameters, variation in disk thickness, disk flaring or the oblateness of bulges will cause correlated deviations across the rotation curves. The effects will be larger at smaller $r$ where higher order multipoles of the potential are more important for the velocity field. My leading-order assumption is meant merely to assess the characteristic impact of uncertainties of this magnitude on the RAR parameters, especially $\sig$.} which are propagated according to
\begin{flalign}
\delta V_\text{bar}^2 = &(0.2 \: \Ug \: V_\text{gas}^2)^2 + (0.2 \: \Ud \: V_\text{disk}^2)^2 \\
&+ (0.2 \: \Ub \: V_\text{bul}^2)^2 \nonumber
\end{flalign}
\vspace{-5mm}
\begin{equation}
\delta \log(\gb) = \sqrt{\log(1+(\delta V_\text{bar}^2/V_\text{bar}^2)^2)}/\ln(10).
\end{equation}
The application of these equations will be indicated by ``boosted uncertainties''.

In this case, the presence of uncertainties in the $x$ direction of the RAR plane introduces latent variables describing the true position of each point on the $x$-axis in the Bayesian hierarchical model. This makes the likelihood function for the parameters of interest alone ambiguous. Two approaches to remove the latent nuisance parameters without sampling them are to marginalise over them with a uniform prior, or to maximise the likelihood with respect to each of them (as a function of the other parameters in the inference) to produce a profile likelihood for the other parameters. These result in different maximum-likelihood points and parameter constraints. Tests on mock data (in agreement with literature results; \citealt{Berger,David}) show that the marginalised likelihood recovers the correct intrinsic scatter and weakly biased shape parameters (e.g. $a_0$ and $\e$) while the profile likelihood recovers unbiased shape parameters but can bias $\sig$ significantly low. As I am mainly interested in whether the boosted uncertainties allow for an intrinsic-scatter-free RAR I opt for the marginalised likelihood, which replaces Eq.~\ref{eq:scatter_1} by
\begin{equation}\label{eq:scatter_2}
\sigma_\text{tot,j}^2 = \delta\log(g_\text{obs,j})^2+\sig^2 + \frac{\text{d}\log\go^\text{pred}}{\text{d}\log\gb}\bigg|_{g_\text{bar,j}}^2 \delta\log(g_\text{bar,j})^2
\end{equation}
in Eq.~\ref{eq:likelihood}
(for the derivation see e.g. sec. 3.2 of \citealt{ESR-RAR}).

I perform the inference using the No U-Turns Sampler (NUTS; \citealt{NUTS}) method of Hamiltonian Monte Carlo (HMC), as implemented in \texttt{numpyro} \citep{phan2019composable, bingham2019pyro}.
I initialise the sampler to the median of 20,000 points randomly drawn from the prior, which I find to yield good convergence behaviour. For each inference I concatenate 28 separate chains run in parallel, manually tuning the number of warmup and sampling steps to ensure that burn-in is complete and that there are enough effective samples for the Gelman-Rubin statistic \citep{Gelman_Rubin} to satisfy $|r-1|<0.001$. This requires $\sim$1000 warmup steps, $\sim$4000 sampling steps and takes $\sim$1 hour per model to run.

\begin{table*}
\begin{center}
\begin{tabular}{|l|c|c|}
  \hline
  \textbf{Parameter} & \textbf{Definition} & \textbf{Prior} \\
  \hline
  \rule{0pt}{3ex}
  $a_0$ & Acceleration constant ($10^{-10}$ ms$^{-2}$) & $\mathcal{U}(0.1,5)$ \\
  \rule{0pt}{3ex}
  $\sig$ & Intrinsic scatter in $\go$ (dex) & $\mathcal{U}(0,1)$ \\
  \rule{0pt}{3ex}
  $\e$ & External field strength relative to $a_0$ & $\mathcal{U}(0,0.5)$ or Lognormal($\log(\bar{e}_\text{N}), \delta\log(\e)$)$^1$\\
  \hline
  \rule{0pt}{3ex}
  $D$ & Distance & $\mathcal{N}\text{t}(\bar{D}, \delta D; 0, \text{None})$ \\
  \rule{0pt}{3ex}
  $i$ & Inclination & $\mathcal{N}\text{t}(\bar{i}, \delta i; 0, 180\deg)$ \\
  \rule{0pt}{3ex}
  $L_{3.6}$ & Luminosity at 3.6$\mu$m & $\mathcal{N}(\bar{L}_{3.6}, \delta L_{3.6})$ \\
  \rule{0pt}{3ex}
  $\Ud$ & $M$/$L$ of disk & Lognormal(-0.72346, 0.24622)$^2$ \\
  \rule{0pt}{3ex}
  $\Ub$ & $M$/$L$ of bulge & Lognormal(-0.38699, 0.24622)$^2$ \\
  \rule{0pt}{3ex}
  $\Ug$ & $M$/$L$ of gas & Lognormal($\log(\bar{\Upsilon}_\text{gas}(M_*))-0.5\log(1.01)$, 0.0997513)$^2$ \\
  \hline
\end{tabular}
\begin{tabular}{c}
$^1$ See Sec.~\ref{sec:model} \quad \quad \quad \quad \quad $^2$ See footnote 2 \\
\end{tabular}
\caption{The free parameters of the inference and their priors. $\mathcal{N}$t denotes a truncated normal with lower and upper limits given by the final two arguments.}
\label{tab:params}
\end{center}
\end{table*}

\subsection{Validation with mock data}
\label{sec:results_mock}

My analysis uses uniform priors on $a_0$, $\sig$ and global $\e$, which are not reparametrisation invariant and cannot, \textit{pace} popular opinion, be considered uninformative.
Such priors are prone to contributing volume effects to the posterior, which can lead to significant biases when applied to parameters to which the likelihood is relatively insensitive (e.g.~\citealt{David}). In addition, the uncertainties and finite sample size lead to scatter in the maximum-likelihood parameters around the population values (sample variance).
To assess the impact of these effects I analyse mock data generated by the following procedure:
\begin{enumerate}[label=(\arabic*),leftmargin=*]
\item Randomly sample the galaxy parameters from their prior distributions
\item Rescale the ``observed'' $\gb$ according to Eq.~\ref{eq:gbar_scaling} to calculate $g_\text{bar}^\text{true}$
\item Use either the no-EFE, global-EFE or max-clustering-EFE model, with some true $a_0$, $\e$, $\sig$, to calculate $g_\text{obs}^\text{true}$ from the $g_\text{bar}^\text{true}$
\item Transform to observed $\go$ through Eq.~\ref{eq:gobs_scaling}, and hence to $V_\text{obs}$ assuming the same $r$ values in the mock data as in the SPARC data
\item Replacing the SPARC $V_\text{obs}$ by these values, calculate the maximum-likelihood values of all parameters and run the inference to compute their posteriors
\item Repeat twice with different random seeds.
\end{enumerate}

Choosing $a_0=1.2$, $\sig=0.05$ dex and $\e=0.01$ in the case with global EFE, the results are shown in Table~\ref{tab:mock_results}. I find neither the maximum-likelihood parameters nor their posteriors to be significantly different to their true values, showing the above effects not to be important for the data and models under consideration. The $\e$ values and galaxy parameters (not shown) are similarly unbiased. This gives confidence to proceed with the analysis of the real data.

\begin{table}
  \begin{center}
    \begin{tabular}{l|c|c|c|c|}
      \hline
       & \multicolumn{2}{c|}{$a_0 \: (1.2)$} & \multicolumn{2}{c|}{$\sig \: (0.05)$}\\ 
\rule{0pt}{4ex}
      \textbf{Model} & ML & \texttt{numpyro} & ML & \texttt{numpyro}\\
      \hline
\rule{0pt}{2ex}
      \multirow{3}{*}{No EFE} & 1.21 & $1.21\pm0.04$ & 0.051 & $0.051\pm0.001$\\
\rule{0pt}{2ex}
      & 1.19 & $1.18\pm0.04$ & 0.049 & $0.049\pm0.001$\\
\rule{0pt}{2ex}
      & 1.21 & $1.15\pm0.04$ & 0.050 & $0.050\pm0.001$\\
\rule{0pt}{2ex}
      \multirow{3}{*}{Global EFE} & 1.20 & $1.29\pm0.05$ & 0.050 & $0.050\pm0.001$\\
\rule{0pt}{2ex}
      & 1.20 & $1.17\pm0.04$ & 0.050 & $0.050\pm0.001$\\
\rule{0pt}{2ex}
      & 1.22 & $1.18\pm0.04$ & 0.050 & $0.049\pm0.001$\\
\rule{0pt}{2ex}
      \multirow{3}{*}{Max-clust EFE} & 1.21 & $1.18\pm0.04$ & 0.050 & $0.051\pm0.001$\\
\rule{0pt}{2ex}
      & 1.18 & $1.14\pm0.04$ & 0.051 & $0.051\pm0.001$\\
\rule{0pt}{2ex}
      & 1.20 & $1.24\pm0.04$ & 0.049 & $0.050\pm0.001$\\
      \hline
  \end{tabular}
  \caption{RAR parameters inferred from mock data generated according to the prescription of Sec.~\ref{sec:results_mock}. ``ML'' stands for maximum likelihood and bracketed numbers after the parameters show the generating values. The HMC uncertainties are $1\sigma$.}
  \label{tab:mock_results}
  \end{center}
\end{table}

\section{Results}
\label{sec:results}

Table~\ref{tab:results} shows the median and $1\sigma$ uncertainty of the RAR parameters for each of the models considered, along with their maximum log-likelihood ($\ln(\hat{L})$), maximum log-posterior ($\ln(\hat{P})$) and Bayesian information criterion (BIC) relative to the first model. The BIC should be taken as a very rough estimator only for the Bayesian evidence, both because the number of data points does not greatly exceed the number of parameters and because the parameter priors are not necessarily slowly varying at the maximum a-posteriori point. For example, replacing $2\ln(\hat{L})$ by $2\ln(\hat{P})$ in the BIC formula changes $\Delta$BIC to 1.44 for the ``No scatter, global EFE'' model, turning ``decisive'' evidence on the Jeffreys scale against the inclusion of $\e$  to ``barely worth mentioning.''
Regardless of this, a clear result from the goodness-of-fit statistics is that either non-zero intrinsic scatter or boosted uncertainties is strongly preferred, mainly through a large increase in the likelihood. The fiducial model, which I considered a priori to be most likely, is ``Scatter, max-clustering EFE''.

\begin{table*}
\begin{center}
\begin{tabular}{|l|c|c|c|c|c|c}
  \hline
  \textbf{Model} & $\mathbf{a_0}$ & $\mathbf{\e}$ & $\mathbf{\sig}$ & $\Delta \ln(\hat{L})$ & $\Delta \ln(\hat{P})$ & $\Delta$BIC\\
  \hline
  \rule{0pt}{4ex}
  No scatter, no EFE & $1.134^{+0.028}_{-0.027}$ & --- & --- & 0 & 0 & 0\\
  \rule{0pt}{4ex}
  No scatter, global EFE & $1.138^{+0.028}_{-0.027}$ & $0.0016^{+0.0005}_{-0.0005}$ & --- & $-3.44$ & 3.24 & 14.8 \\
  \rule{0pt}{4ex}
  No scatter, max-clustering EFE & $1.309^{+0.039}_{-0.037}$ & $0.0050^{+0.0203}_{-0.0033}$ & --- & 292 & 55.8 & 577 \\
  \rule{0pt}{4ex}
  No scatter, avg-clustering EFE & $1.307^{+0.038}_{-0.036}$ & $0.0022^{+0.0231}_{-0.0017}$ & --- & 275 & $-19.2$ & 611 \\
  \rule{0pt}{4ex}
  Scatter, no EFE & $1.070^{+0.032}_{-0.031}$ & --- & $0.035^{+0.001}_{-0.001}$ & 2020 & 2381 & $-4032$ \\
  \rule{0pt}{4ex}
  As above, boosted uncertainties & $1.121^{+0.035}_{-0.034}$ & --- & $<0.0020$ & 1936 & 2354 & $-3864$ \\
  \rule{0pt}{4ex}
  Scatter, global EFE & $1.077^{+0.033}_{-0.032}$ & $0.0017^{+0.0009}_{-0.001}$ & $0.035^{+0.001}_{-0.001}$ & 2019 & 2375 & $-4023$\\
  \rule{0pt}{4ex}
  Scatter, max-clustering EFE & $1.236^{+0.043}_{-0.041}$ & $0.0049^{+0.0125}_{-0.003}$ & $0.033^{+0.001}_{-0.001}$ & 2052 & 2220 & $-2935$ \\
  \rule{0pt}{4ex}
  As above, boosted uncertainties & $1.275^{+0.047}_{-0.044}$ & $0.0047^{+0.0103}_{-0.0029}$ & $<0.0019$ & 1948 & 2182 & $-2727$ \\
  \rule{0pt}{4ex}
  Scatter, avg-clustering EFE & $1.272^{+0.047}_{-0.045}$ & $0.0020^{+0.0134}_{-0.0016}$ & $0.033^{+0.001}_{-0.001}$ & 2074 & 2161 & $-2978$ \\
  \hline
\end{tabular}
\caption{Constraints on RAR parameters and goodness-of-fit statistics for the models considered. For the maximum-clustering and average-clustering EFE models the quoted $\e$ constraints describe the stacked posteriors over all galaxies. The final three columns are the maximum log-likelihood, maximum log-posterior and Bayesian information criterion relative to the first model.}
\label{tab:results}
\end{center}
\end{table*}

The value of $a_0$ is slightly reduced if the RAR is assumed to possess intrinsic scatter, and more significantly increased by including the EFE with external field strength priors from the baryonic large-scale structure. This is because these priors somewhat increase $\e$ relative to likelihood alone and there is a positive degeneracy between $\e$ and $a_0$. Differences in $a_0$ between the models are up to a few times larger than their statistical uncertainties, and a
naive averaging over all the models implies $a_0=1.19 \pm 0.04 \, \text{(stat)} \pm 0.09 \, \text{(sys)}$.

The intrinsic scatter is similar in all models in which it is included, except in the ``boosted uncertainties'' case. The results suggest that $\sig=0.034 \pm 0.001 \, \text{(stat)} \pm 0.001 \, \text{(sys)}$ dex, with the significant caveat that this assumes the SPARC error model is reliable. A 10 per cent uncertainty on $V_\text{disk}$, $V_\text{bul}$ and $V_\text{disk}$---as may be expected from deviations from the assumed 3D baryon geometry---is sufficient to set the preferred $\sig$ to 0. The intrinsic scatter is not driven by the most egregious outliers: removing the 11 points with $\gb>\go$ and $\go<2$ after transformation for the model without EFE (see Fig.~\ref{fig:transformed}) reduces it only to 0.031 dex.

There is weak evidence for $\langle\e\rangle>0$ when it is given a uniform prior, as evidenced both by the inferred $\e$ being consistent with 0 within $\sim$3$\sigma$ and by the small gain in maximum posterior value across the chain ($\Delta \ln(\hat{P})$). Adding this parameter is not favoured by the BIC. The model perhaps most similar to the MOND expectation with global $\e$ (not shown in Table~\ref{tab:results}) has $\sig=0$ and boosted uncertainties: in this case $\langle\e\rangle=0.0024\pm0.0009$, again a 3$\sigma$ ``detection'' but a slightly larger value. This model has $\Delta$BIC$=-3860$, very similar to the case with $\sig$ and boosted uncertainties but without EFE, showing that again the addition of a global $\e$ is not favoured.

The average prior values of $\e$ over all the galaxies for the maximum-clustering and average-clustering cases are 0.0050 and 0.0018 respectively, while the posteriors average to 0.0050 and 0.0022 without $\sig$ and 0.0049 and 0.0020 with. This shows that when modelled galaxy-by-galaxy the data does not disfavour significant values of $\e$ in agreement with the large-scale structure expectation, and in fact in both cases the maximum-likelihood value is increased over the case of no or global $\e$, significantly without intrinsic scatter and moderately with it. Prior evidence is however required to support the presence of the EFE in a model comparison sense as the BIC strongly disfavours the addition of 147 $\e$ parameters. While the average-clustering EFE model gives a slightly higher $\hat{L}$ when including $\sig$, it gives a lower $\hat{L}$ without it and a lower $\hat{P}$ in both cases. This constitutes weak evidence in favour of the maximum-clustering prior, but the data is far from sufficient to distinguish robustly between them.

Fig.~\ref{fig:transformed} uses Eqs.~\ref{eq:gbar_scaling} and~\ref{eq:gobs_scaling} to transform $\gb$ and $\go$ according to the parameters at the median of the posterior for the inference with $\sig$ but without EFE (the model preferred by the BIC), including the $2\sigma$ model prediction. This illustrates the extreme tightness of the underlying (posterior) RAR, including relative to the traditional ``prior RAR'' shown in grey. The few outlying points are also outliers of the prior RAR, and are brought slightly closer to the line by the transformation. Note that the systematic uncertainty on the blue points due to their dependence on the galaxy nuisance parameters is not shown.

\begin{figure}
  \centering
  \includegraphics[width=0.5\textwidth]{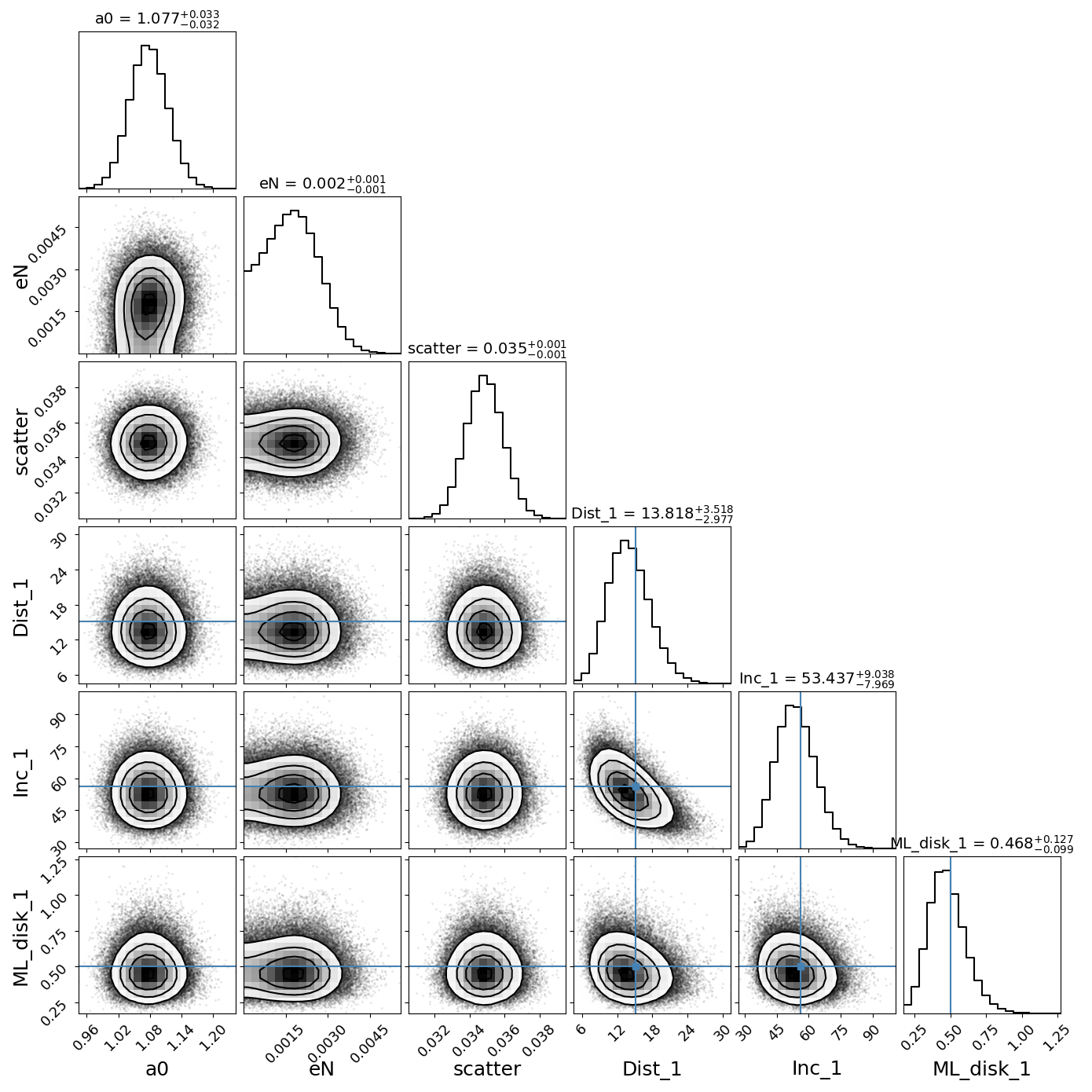}
  \caption{Corner plot of selected parameters from the inference including intrinsic scatter and global $\e$. For the distance (in Mpc), inclination (in degrees) and disk mass-to-light ratio of the first galaxy (D512-2), the maximum a priori values are shown by the blue lines.}
  \label{fig:corner}
\end{figure}

Fig.~\ref{fig:corner} shows an excerpt from the corner plot of the inference that also includes the EFE with a global $\e$. The posteriors on three parameters for galaxy 1 (D512-2) are compared to their maximum prior values (blue lines), with which they agree well. $a_0$, $\e$ and $\sig$ are not strongly degenerate with other parameters, while
the full parameter space exhibits the degeneracies expected from Eqs.~\ref{eq:gbar_scaling} and~\ref{eq:gobs_scaling}.
The median distance to the SPARC galaxies is reduced by 1.7 per cent going from the inference with $\sig$ but without EFE to that with local $\e$ and maximum-clustering prior, although within the latter inference $D$ is positively correlated with $\e$ on average within the chain. This is because at higher $\e$, $\go^\text{pred}$ is lower at low $\gb$ and hence closer to $\go$ at higher $D$ (or $i$).

\begin{figure}
  \centering
  \includegraphics[width=0.46\textwidth]{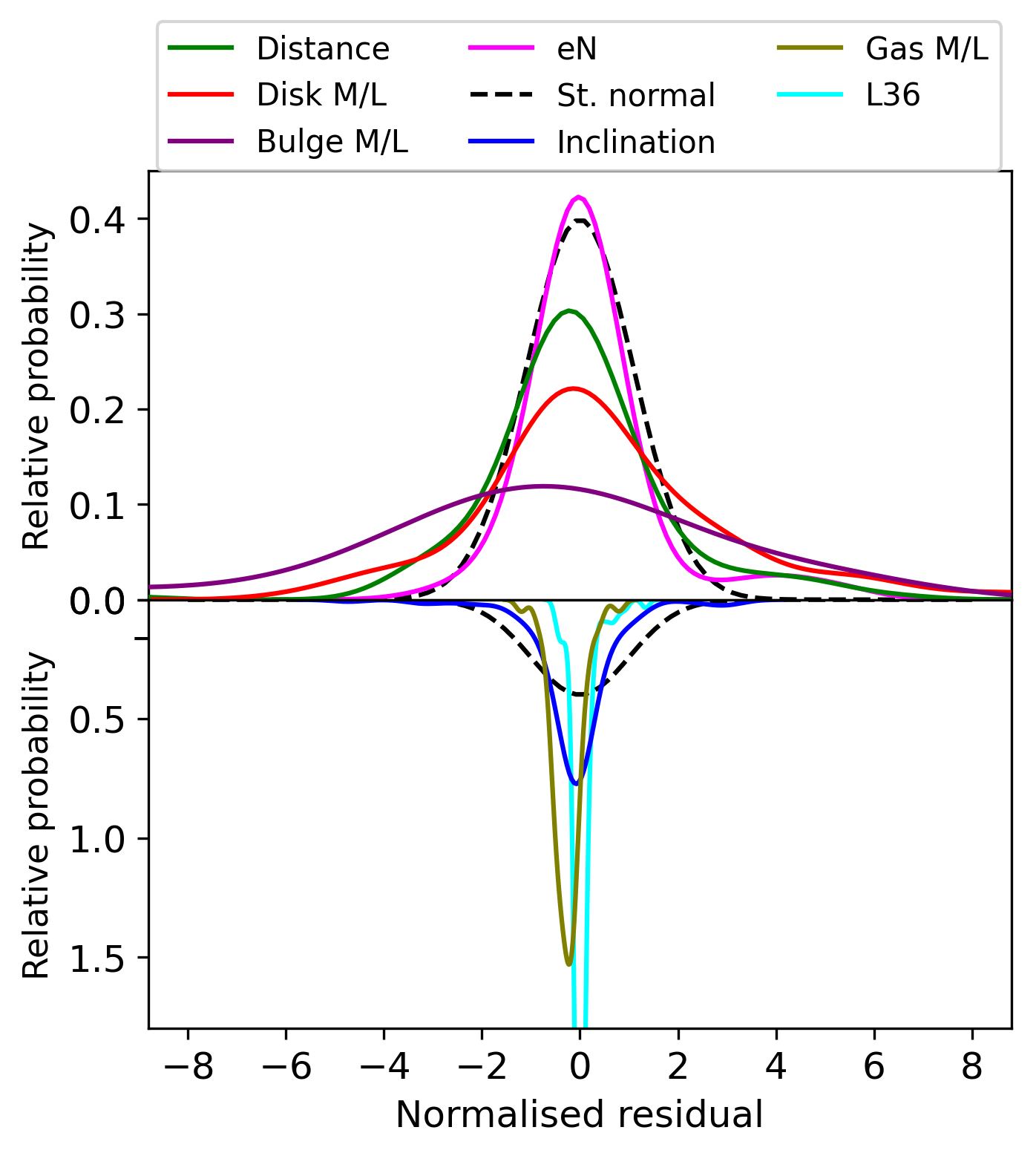}
  \caption{The distribution of normalised residuals (Eq.~\ref{eq:norm_res}) of the galaxy-specific parameters in the inference including intrinsic scatter and local $\e$ with maximum-clustering prior. The mirroring of some of the curves about the $x$-axis is for visual clarity only. A standard normal distribution---corresponding to $X_\text{post}$ scattering around $\bar{X}$ as a Gaussian with width given by std($X_\text{post}$)---is shown in dashed black.}
  \label{fig:residuals}
\end{figure}

To show more generally the differences between the prior and posterior values of the galaxy parameters, Fig.~\ref{fig:residuals} shows the distribution of normalised residuals for the model with intrinsic scatter and galaxy-by-galaxy $\e$ with maximum-clustering prior. For galaxy parameter $X$, the normalised residual is defined as
\begin{equation}\label{eq:norm_res}
(\text{med}(X_\text{post})-\bar{X})/\text{std}(X_\text{post})
\end{equation}
where subscript ``post'' and ``prior'' denote the posterior and prior distributions, and ``med'' and ``std'' stand for median and standard deviation. Larger values of the residual indicate a posterior significantly shifted from the prior due to the influence of the likelihood. The width of the distributions therefore reflect the sensitivity of the data to the parameters: those with distributions sharply peaked at 0 such as $L_{3.6}$ and $\Ug$ are relatively unimportant so that $\text{med}({X}_\text{post})\approx\bar{X}$, while those with very broad distributions have likelihood often peaked far from the prior centres. This is especially pronounced for $\Ub$ and $\Ud$ which determine $\gb$ in the galaxies' central regions. $\Ub$ in particular has a slight negative offset to reduce the number of inner points for which $\gb>\go$. The black dashed lines are what one would expect if $X_\text{post}$ scatters around $\bar{X}$ with a Gaussian distribution of width given by std($X_\text{post}$).

\begin{figure}
  \centering
  \includegraphics[width=0.5\textwidth]{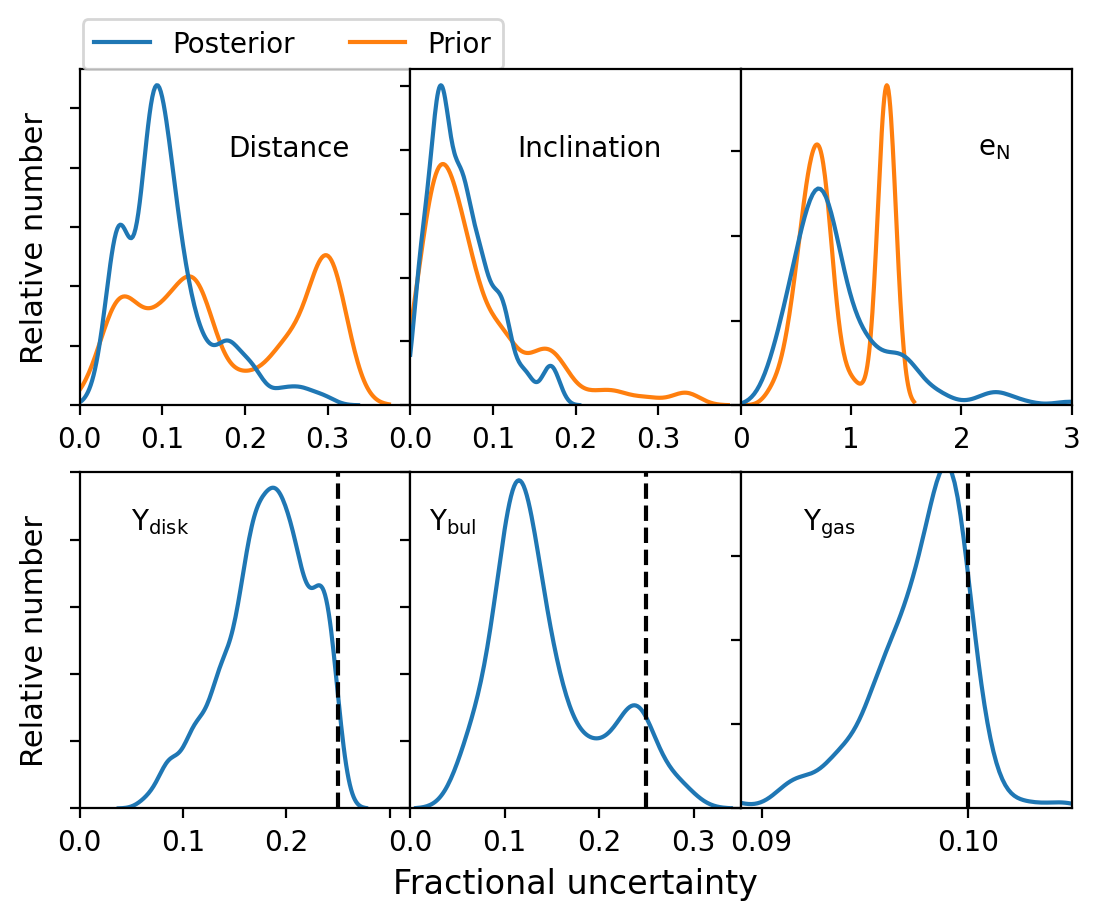}
  \caption{Distribution of prior and posterior fractional uncertainties on galaxy-specific parameters for the inference with scatter and local $\e$ with maximum-clustering prior. In the lower panels the vertical dashed lines are the fractional prior uncertainties (25 and 10 per cent), which are the same for all galaxies.}
  \label{fig:prior-post}
\end{figure}

Finally, I show in Fig.~\ref{fig:prior-post} smoothed distributions of the fractional uncertainties in the galaxy-specific parameters in the prior and posterior for the inference with intrinsic scatter and local $\e$ with maximum-clustering prior. In most cases the RAR constraint has increased the precision with which the parameters are known. This is especially marked for the distance, where galaxies in the second mode of the prior distribution (those at $\lesssim$60 Mpc with only redshift distances; \citealt{SPARC}) are brought into a single posterior mode at $\sim$10 per cent uncertainty. This illustrates the utility of the RAR as a direct (i.e. redshift-independent) distance measurement method. There is analogous behaviour for $\e$, where the higher prior mode corresponds to galaxies outside the SDSS footprint which are assigned higher uncertainties (see Sec.~\ref{sec:model}); these become better known on applying the RAR constraint. The uncertainties on $\Ud$ and $\Ub$ also fall markedly, partly to reduce the number of inner points with posterior probability at $\gb>\go$.\footnote{An alternative explanation for apparent $\gb>\go$ at low $r$ is the presence of bars or asymmetries in the inner regions of the disk. In some cases it may be preferable to excise such data (e.g.~\citealt{Harley_asym}) to prevent it from biasing $\Ud$ low, which has a knock-on effect across the rotation curve.} The analogues of Figs.~\ref{fig:transformed}--\ref{fig:prior-post} for the other models are qualitatively similar, in line with the variation in their results shown in Table~\ref{tab:results}.

Supplementary tables available at \url{https://zenodo.org/record/7752545} \citep{zenodo} contain the mean, median and 1, 2 and 3$\sigma$ constraints on the parameters for each of the models considered. The format is illustrated in Table~\ref{tab:supplemental}, which shows the first five and last two rows for the model with $\sig$ and global $\e$. An example use of these is to resolve distance variations among galaxies in the Ursa Major cluster, all of which have $\bar{D}=18$ Mpc. A reference table contains the galaxy names at each index as well as the bulge luminosity and means and standard deviations of their Gaussian priors on $D$, $i$ and $L_{3.6}$, transcribed from the SPARC database.
Plots of the prior and posterior constraints on $D$, $i$, $L_{3.6}$, $\e$, $\Ud$, $\Ub$ and $\Ug$ for all galaxies under each model are also included.

\begin{table*}
\begin{center}
\begin{tabular}{|l|c|c|c|c|c|c|c|c|}
  \hline
  \textbf{Parameter} & 0.135\% & 2.275\% & 15.87\% & 50\% & mean & 84.13\% & 97.725\% & 99.865\%\\
  \hline
  \rule{0pt}{2ex}
  $a_0$ & 0.9847 & 1.0145 & 1.0455 & 1.0773 & 1.078 & 1.1102 & 1.145 & 1.1792\\
  \rule{0pt}{2ex}
  $\sig$ & 0.0321 & 0.033 & 0.0339 & 0.0349 & 0.0349 & 0.0358 & 0.0368 & 0.0379\\
  \rule{0pt}{2ex}
  $\e$ & 0.0 & 0.0001 & 0.0007 & 0.0017 & 0.0017 & 0.0026 & 0.0035 & 0.0044\\
  \rule{0pt}{2ex}
  Dist[0] & 2.0319 & 2.2868 & 2.5538 & 2.8242 & 2.8252 & 3.0978 & 3.3688 & 3.6426\\
  \rule{0pt}{2ex}
  Dist[1] & 6.6289 & 8.4419 & 10.8253 & 13.8181 & 14.0917 & 17.3551 & 21.3172 & 25.6289\\
  \rule{0pt}{2ex}
  \vdots  & \vdots & \vdots & \vdots & \vdots & \vdots  & \vdots & \vdots & \vdots \\
  \rule{0pt}{2ex}
  ML\_gas[145] & 0.6955 & 0.7658 & 0.8427 & 0.9271 & 0.9308 & 1.0185 & 1.1198 & 1.2295\\
  \rule{0pt}{2ex}
  ML\_gas[146] & 0.7595 & 0.832 & 0.9139 & 1.0046 & 1.009 & 1.1038 & 1.2145 & 1.3344\\
  \hline
\end{tabular}
\caption{Excerpt of posterior summaries for the model with intrinsic scatter and global $\e$. The full tables for all models are available online.}
\label{tab:supplemental}
\end{center}
\end{table*}

\section{Discussion}
\label{sec:disc}


While the value of $a_0$ that I infer ($1.1 \lesssim a_0 \lesssim 1.3$) is in full agreement with literature results, the RAR intrinsic scatter $\sig\approx0.034$ dex is significantly smaller. \citet{RAR} quote a total scatter of 0.13 dex and argue that most of this comes from observational uncertainties; using their model and Eqs.~\ref{eq:likelihood} and~\ref{eq:scatter_2} I find $\sig=0.082\pm0.003$ dex.
\citet{Li} quote an intrinsic scatter of 0.057 dex from residuals around their best-fit relation with fixed $a_0=1.2$.
A reduction in intrinsic scatter when marginalising over galaxy parameters is not guaranteed: although the denominator of Eq.~\ref{eq:likelihood} favours smaller $\sig$, this is offset by lower prior probabilities of the galaxy parameters at values that bring the points closer to the theoretical line to reduce the exponent in Eq.~\ref{eq:likelihood}. The statistical uncertainties on $\gb$ and $\go$ also fall greatly (the former to 0) when the galaxy parameters are inferred, so a lower total scatter of the points around the line need not translate into lower $\sig$. That the preferred intrinsic scatter in the full analysis is extremely small hints towards the RAR being at base a practically monotonic correlation.

The question of the fundamentality of the RAR has important ramifications for the mass discrepancy problem on galaxy scales. \citet{RAR} claimed the relation to be law-like, a result supported here. It will be challenging for simulations or semi-analytic models in $\Lambda$CDM to reproduce the tightness of the underlying RAR given that even the scatter of the prior RAR, $\sim$0.1 dex, is non-trivial~\citep{DC_Lelli,Desmond_MDAR,Keller,Ludlow}. It would be interesting to explore this explicitly using modern high-resolution cosmological hydrodynamical $\Lambda$CDM simulations such as TNG-50~\citep{TNG50_1,TNG50_2} and NewHorizon~\citep{NH}, which would provide a stringent test of those models.

It is important to bear in mind that SPARC is only a small fraction of the data pertinent to the RAR: one may also use ultra-diffuse galaxies (e.g.~\citealt{Freundlich}), local dwarf spheroidals (e.g.~\citealt{McGaugh_Wolf, McGaugh_Milgrom}), early-type galaxies (e.g.~\citealt{Yu_early, ellipticals_1}), low-acceleration regions including the outer Milky Way \citep{Oman}, stacked weak lensing \citep{Brouwer} and groups and clusters of galaxies \citep{ellipticals_2, ellipticals_1, clusters_1, clusters_2, clusters_3, groups}. Some of this data appears to deviate from the MOND expectation. SPARC is however one of the few datasets with uncertainties under sufficient control for the present analysis to be feasible and meaningful. Future work may extend it to new regimes.

Na\"ively one expects $\sig=0$ in MOND, but in fact this is only true in modified inertia formulations in the limit of perfectly circular orbits. Even in such models a scatter is introduced by deviations from circularity because the dynamics of an object depends on its entire past trajectory~\citep{inertia_2,inertia_1}. This may already be sufficient to account for the 0.034 dex (8 per cent) scatter present in the underlying RAR, suggesting that modified inertia, which predicts most directly the algebraic relation that I fit, is viable. Additional scatter is present in modified gravity formulations where the algebraic MOND relation holds only in spherical symmetry~\citep{Famaey_McGaugh}, a condition clearly violated in disk galaxies. It would be interesting to quantify the scatter introduced by these effects in SPARC-like galaxies, further testing the MOND paradigm and providing a novel way to distinguish between the modified gravity and modified inertia interpretations~\citep{Peterson,Chae_distinguishing}. The model of constant mass-to-light for the disk and bulge may also be overly simplistic, with radial dependence parameters able to soak up some of the remaining scatter.

The presence or absence of the EFE is controversial within the MOND literature, with some studies claiming strong evidence for it~\citep{McGaugh_Milgrom,Haghi,Paper_I} and others strong evidence against~\citep{Hernandez_Cookson_Cortes,Hernandez_Lara,Freundlich}. My work does not resolve this issue: there is weak evidence for a positive average external field strength across the sample, while inferring it galaxy-by-galaxy with a prior from independent measurements of environment improves the likelihood but is not favoured by the Bayesian information criterion. An important caveat is that the fitting formula I use (Eq.~\ref{eq:efe}) was only designed for the outer regions of rotation curves~\citep{Chae_Milgrom}---while I am applying it to them in their entirety---and is only valid within the AQUAL model. Different EFE formulae may give different constraints on $\e$, and hence $a_0$ with which it is degenerate, but would not be expected to alter $\sig$ appreciably. (A scatter in the effect of the EFE, e.g. due to variably internal and external fields, may however reduce the true intrinsic scatter.) The environmental priors are also highly uncertain due to the possibility of clustered unseen baryonic mass. Further work on modelling the EFE and constraining the external field is therefore required to reach a definitive conclusion concerning the existence of the EFE in galaxy dynamics. As MOND is currently an effective model only, it may be that the underlying theory gives a mass or scale dependence to the EFE which can reconcile seemingly discrepant results.

Besides calibrating the RAR my work provides strong constraints on the properties of the SPARC galaxies under the assumption that the underlying RAR is as I model it. Summaries of these constraints are made public in online tables to facilitate future studies using SPARC. One such application is to use the RAR as a direct distance probe. The current study calibrates the relation using a sample with informative distance priors. The distance to any galaxy with a (partially) resolved rotation curve may be inferred by fitting it to the calibrated RAR, including marginalisation over the other relevant properties of the galaxy but not necessarily the parameters of the RAR itself. This is analogous to the well-established Tully--Fisher and Fundamental Plane methods, but achieves higher precision ($\sim$10 per cent uncertainty on $D$ rather than 20-25 per cent; \citealt{CF4}) at the cost of requiring resolved kinematics. This may readily be achieved for a large sample of galaxies using the high spatial resolution of upcoming instruments such as the Square Kilometer Array. Note however that the RAR may evolve with redshift, e.g. due to time-dependent $a_0$ in MOND or evolution of galaxy and halo density profiles in $\Lambda$CDM~\citep{Keller, Paranjape_Sheth}, which would necessitate recalibration of the relation when this effect kicks in. My analysis also supplies enhanced kinematic inclinations, as well as constraints on mass-to-light ratios which may be correlated with other galaxy properties to advance understanding of stellar populations and galaxies' gas content.

Outliers of the underlying RAR may either be individual rotation curve points with large residuals from the best-fit line (e.g. in Fig.~\ref{fig:transformed}) or entire galaxies with parameters strongly shifted from their prior centres to achieve a good fit (readily visible in the supplementary figures). Studying these on a case-by-case basis may help identify peculiar galactic features and signpost the need for more sophisticated modelling. For example, NGC 2915 seems to require very high $\Ud\approx1.2$ in all models. This is a starburst dwarf galaxy with significant radial motion in the inner regions, the modelling of which affects the entire rotation curve. It has a highly complex structure and is likely not in dynamical equilibrium towards the centre, while further out there is a strong warp~\citep{Meurer,Elson_3,Elson_1,Elson_2,Tang}. The ability to correlate such properties with the results of this analysis across the sample would lend weight to the interpretation of the underlying RAR as fundamental.

Provided the fitting functions are good, the constraints on galaxy parameters do not assume MOND any more than modelling the Tully--Fisher relation as a power-law. Both may be thought of as empirical descriptions of the data without consideration of their theoretical significance.
Recently, however, \citet{ESR-RAR} have challenged the optimality of MOND functions (those with Newtonian and deep-MOND or EFE-driven regimes) for fitting the RAR, finding that the majority of functions that most efficiently compress the SPARC data tend to constant $\go$ at low $\gb$. These functions and their parameters have no known theoretical significance, but would lead to different ``underlying'' relations and hence galaxy parameter constraints. This systematic uncertainty in e.g. distance measurements could be assessed by repeating the present inference with one or more of these functions.
It is however not known how the results of \citeauthor{ESR-RAR} would be affected by marginalising over the galaxy parameters separately for each function as done here, which would alter the function ranking, or by extending Exhaustive Symbolic Regression \citep{ESR} to higher complexity.

I see three technical ways in which this inference could be improved. The first is to separate $\delta V_\text{obs}$ into a truly statistical and systematic part, including its full covariance across the galaxies' rotation curves. The average $\delta \log(V_\text{obs})$ across the sample is 0.03 dex, so more accurate modelling of this has the potential to alter the best-fit value of $\sig$, which is similar. In particular, an important contribution to $\delta V_\text{obs}$ is from the difference in velocity between the approaching and receding sides of the disk, which is likely strongly correlated over $r$. This may itself be correlated with inclination, which in detail may vary across the disk. The second is to characterise better the molecular gas and 3D baryon geometry. The former could be done by estimating (or sampling) $M_{H_2}$ from the $M_*-M_{H_2}$ relation of a population of similar galaxies and adding a contribution to $V_\text{bar}$ from a thin $H_2$ disk, and the latter by solving Poisson's equation for different assumptions about disk thickness, bulge oblateness and potential asymmetries. For the SPARC galaxies these uncertainties are however highly subdominant to those that I model explicitly. The third is to use a Jeffreys rather than uniform prior for $a_0$, $\e$ and $\sig$, thus eliminating potential volume effects. The mock tests of Sec.~\ref{sec:results_mock} show these not to bias the results significantly. A Jeffreys prior would however enable the inference of $\e$ galaxy-by-galaxy without importing large-scale structure information; when I tried this on mock data using a uniform prior I found $a_0$ to be biased high, presumably due to the allowed and poorly-constrained volume towards high $\e$. Another extension would be to try a more complex intrinsic scatter model; \citet{Li} for example find a superposition of two Gaussians to fit the residuals better than one, perhaps reflecting the separate formal error and kinematic asymmetry contributions to $V_\text{obs}$.


\section{Conclusion}
\label{sec:conc}

I have uncovered underlying RARs in the SPARC data by fitting the parameters of the Simple interpolating function, with and without intrinsic scatter and the external field effect, simultaneously with all relevant galaxy properties. The preferred intrinsic scatter is very small, $0.034\pm0.002$ dex, and additional plausible uncertainties are capable of reducing this to 0. The acceleration constant is in the range $1.1 \lesssim a_0 \lesssim 1.3$, in good agreement with literature results.
I find weak evidence for the external field effect using an average external field strength over the full sample with a uniform prior. Allowing the field strength to vary galaxy-by-galaxy with a prior from large-scale structure observations improves the overall likelihood of the data but is not favoured by the Bayesian information criterion. My results suggest near-monotinicity and a high degree of regularity in the the RAR, providing a fresh challenge to galaxy formation models. The constraints I produce on the SPARC galaxies' parameters---distance, inclination, luminosity and disk, bulge and gas mass-to-light ratios---are the most precise to date (although subject to systematic error if the underlying RAR is not as I model it). I publicly release summaries of all posteriors for analyses that may benefit from this information.

\section{Data availability}

The mean, median and 1, 2 and 3$\sigma$ confidence intervals of the parameters for all models are available at \url{https://zenodo.org/record/7752545}.
The remaining data generated here, including the full HMC chains to explore degeneracies, will be made available on reasonable request.

\section*{Acknowledgements}

I thank Indranil Banik, Deaglan Bartlett, Kyu-Hyun Chae, Benoit Famaey, Pedro Ferreira, Xavier Hernandez, Federico Lelli, Stacy McGaugh, Mordehai Milgrom, James Prideaux-Ghee, Richard Stiskalek and Tariq Yasin for useful discussions.

I am supported by a Royal Society University Research Fellowship (grant no. 211046). This project has received funding from the European Research Council (ERC) under the European Union's Horizon 2020 research and innovation programme (grant agreement No 693024).

For the purpose of open access, I have applied a Creative Commons Attribution (CC BY) licence to any Author Accepted Manuscript version arising.

\bibliographystyle{mnras}
\bibliography{references}

\begin{thebibliography}{}
\makeatletter
\relax
\def\mn@urlcharsother{\let\do\@makeother \do\$\do\&\do\#\do\^\do\_\do\%\do\~}
\def\mn@doi{\begingroup\mn@urlcharsother \@ifnextchar [ {\mn@doi@}
  {\mn@doi@[]}}
\def\mn@doi@[#1]#2{\def\@tempa{#1}\ifx\@tempa\@empty \href
  {http://dx.doi.org/#2} {doi:#2}\else \href {http://dx.doi.org/#2} {#1}\fi
  \endgroup}
\def\mn@eprint#1#2{\mn@eprint@#1:#2::\@nil}
\def\mn@eprint@arXiv#1{\href {http://arxiv.org/abs/#1} {{\tt arXiv:#1}}}
\def\mn@eprint@dblp#1{\href {http://dblp.uni-trier.de/rec/bibtex/#1.xml}
  {dblp:#1}}
\def\mn@eprint@#1:#2:#3:#4\@nil{\def\@tempa {#1}\def\@tempb {#2}\def\@tempc
  {#3}\ifx \@tempc \@empty \let \@tempc \@tempb \let \@tempb \@tempa \fi \ifx
  \@tempb \@empty \def\@tempb {arXiv}\fi \@ifundefined
  {mn@eprint@\@tempb}{\@tempb:\@tempc}{\expandafter \expandafter \csname
  mn@eprint@\@tempb\endcsname \expandafter{\@tempc}}}

\bibitem[\protect\citeauthoryear{{Banik} \& {Zhao}}{{Banik} \&
  {Zhao}}{2015}]{Banik_EFE}
{Banik} I.,  {Zhao} H.,  2015, \mn@doi [arXiv e-prints]
  {10.48550/arXiv.1509.08457}, \href
  {https://ui.adsabs.harvard.edu/abs/2015arXiv150908457B} {p. arXiv:1509.08457}

\bibitem[\protect\citeauthoryear{{Banik} \& {Zhao}}{{Banik} \&
  {Zhao}}{2022}]{Banik}
{Banik} I.,  {Zhao} H.,  2022, \mn@doi [Symmetry] {10.3390/sym14071331}, \href
  {https://ui.adsabs.harvard.edu/abs/2022Symm...14.1331B} {14, 1331}

\bibitem[\protect\citeauthoryear{{Bartlett}, {Desmond}  \&
  {Ferreira}}{{Bartlett} et~al.}{2022}]{ESR}
{Bartlett} D.~J.,  {Desmond} H.,   {Ferreira} P.~G.,  2022, arXiv e-prints,
  \href {https://ui.adsabs.harvard.edu/abs/2022arXiv221111461B} {p.
  arXiv:2211.11461}

\bibitem[\protect\citeauthoryear{{Bekenstein} \& {Milgrom}}{{Bekenstein} \&
  {Milgrom}}{1984}]{AQUAL}
{Bekenstein} J.,  {Milgrom} M.,  1984, \mn@doi [\apj] {10.1086/162570}, \href
  {https://ui.adsabs.harvard.edu/abs/1984ApJ...286....7B} {286, 7}

\bibitem[\protect\citeauthoryear{Berger, Liseo  \& Wolpert}{Berger
  et~al.}{1999}]{Berger}
Berger J.~O.,  Liseo B.,   Wolpert R.~L.,  1999, \mn@doi [Statistical Science]
  {10.1214/ss/1009211804}, 14, 1

\bibitem[\protect\citeauthoryear{Bingham et~al.,}{Bingham
  et~al.}{2019}]{bingham2019pyro}
Bingham E.,  et~al., 2019, J. Mach. Learn. Res., 20, 28:1

\bibitem[\protect\citeauthoryear{{Blanton}, {Geha}  \& {West}}{{Blanton}
  et~al.}{2008}]{Blanton}
{Blanton} M.~R.,  {Geha} M.,   {West} A.~A.,  2008, \mn@doi [\apj]
  {10.1086/588800}, \href
  {https://ui.adsabs.harvard.edu/abs/2008ApJ...682..861B} {682, 861}

\bibitem[\protect\citeauthoryear{{Brouwer} et~al.,}{{Brouwer}
  et~al.}{2021}]{Brouwer}
{Brouwer} M.~M.,  et~al., 2021, \mn@doi [\aap] {10.1051/0004-6361/202040108},
  \href {https://ui.adsabs.harvard.edu/abs/2021A&A...650A.113B} {650, A113}

\bibitem[\protect\citeauthoryear{{Cappellari} et~al.,}{{Cappellari}
  et~al.}{2013}]{Cappellari}
{Cappellari} M.,  et~al., 2013, \mn@doi [\mnras] {10.1093/mnras/stt562}, \href
  {https://ui.adsabs.harvard.edu/abs/2013MNRAS.432.1709C} {432, 1709}

\bibitem[\protect\citeauthoryear{Chae}{Chae}{2022}]{Chae_distinguishing}
Chae K.-H.,  2022, \mn@doi [\apj] {10.3847/1538-4357/ac93fc}, 941, 55

\bibitem[\protect\citeauthoryear{{Chae} \& {Milgrom}}{{Chae} \&
  {Milgrom}}{2022}]{Chae_Milgrom}
{Chae} K.-H.,  {Milgrom} M.,  2022, \mn@doi [\apj] {10.3847/1538-4357/ac5405},
  \href {https://ui.adsabs.harvard.edu/abs/2022ApJ...928...24C} {928, 24}

\bibitem[\protect\citeauthoryear{{Chae}, {Bernardi}, {Sheth}  \& {Gong}}{{Chae}
  et~al.}{2019}]{ellipticals_2}
{Chae} K.-H.,  {Bernardi} M.,  {Sheth} R.~K.,   {Gong} I.-T.,  2019, \mn@doi
  [\apj] {10.3847/1538-4357/ab18f8}, \href
  {https://ui.adsabs.harvard.edu/abs/2019ApJ...877...18C} {877, 18}

\bibitem[\protect\citeauthoryear{{Chae}, {Bernardi}, {Dom{\'\i}nguez
  S{\'a}nchez}  \& {Sheth}}{{Chae} et~al.}{2020a}]{ellipticals_1}
{Chae} K.-H.,  {Bernardi} M.,  {Dom{\'\i}nguez S{\'a}nchez} H.,   {Sheth}
  R.~K.,  2020a, \mn@doi [\apjl] {10.3847/2041-8213/abc2d3}, \href
  {https://ui.adsabs.harvard.edu/abs/2020ApJ...903L..31C} {903, L31}

\bibitem[\protect\citeauthoryear{{Chae}, {Lelli}, {Desmond}, {McGaugh}, {Li}
  \& {Schombert}}{{Chae} et~al.}{2020b}]{Paper_I}
{Chae} K.-H.,  {Lelli} F.,  {Desmond} H.,  {McGaugh} S.~S.,  {Li} P.,
  {Schombert} J.~M.,  2020b, \mn@doi [\apj] {10.3847/1538-4357/abbb96}, \href
  {https://ui.adsabs.harvard.edu/abs/2020ApJ...904...51C} {904, 51}

\bibitem[\protect\citeauthoryear{{Chae}, {Desmond}, {Lelli}, {McGaugh}  \&
  {Schombert}}{{Chae} et~al.}{2021}]{Paper_II}
{Chae} K.-H.,  {Desmond} H.,  {Lelli} F.,  {McGaugh} S.~S.,   {Schombert}
  J.~M.,  2021, \mn@doi [\apj] {10.3847/1538-4357/ac1bba}, \href
  {https://ui.adsabs.harvard.edu/abs/2021ApJ...921..104C} {921, 104}

\bibitem[\protect\citeauthoryear{{Chae}, {Lelli}, {Desmond}, {McGaugh}  \&
  {Schombert}}{{Chae} et~al.}{2022}]{Paper_III}
{Chae} K.-H.,  {Lelli} F.,  {Desmond} H.,  {McGaugh} S.~S.,   {Schombert}
  J.~M.,  2022, \mn@doi [\prd] {10.1103/PhysRevD.106.103025}, \href
  {https://ui.adsabs.harvard.edu/abs/2022PhRvD.106j3025C} {106, 103025}

\bibitem[\protect\citeauthoryear{{Chan} \& {Del Popolo}}{{Chan} \& {Del
  Popolo}}{2020}]{clusters_1}
{Chan} M.~H.,  {Del Popolo} A.,  2020, \mn@doi [\mnras]
  {10.1093/mnras/staa225}, \href
  {https://ui.adsabs.harvard.edu/abs/2020MNRAS.492.5865C} {492, 5865}

\bibitem[\protect\citeauthoryear{{Desmond}}{{Desmond}}{2017}]{Desmond_MDAR}
{Desmond} H.,  2017, \mn@doi [\mnras] {10.1093/mnras/stw2571}, \href
  {http://adsabs.harvard.edu/abs/2017MNRAS.464.4160D} {464, 4160}

\bibitem[\protect\citeauthoryear{Desmond}{Desmond}{2023}]{zenodo}
Desmond H.,  2023, \mn@doi{10.5281/zenodo.7752545}

\bibitem[\protect\citeauthoryear{{Desmond} \& {Wechsler}}{{Desmond} \&
  {Wechsler}}{2015}]{Desmond_TFR}
{Desmond} H.,  {Wechsler} R.~H.,  2015, \mn@doi [\mnras]
  {10.1093/mnras/stv1978}, \href
  {https://ui.adsabs.harvard.edu/abs/2015MNRAS.454..322D} {454, 322}

\bibitem[\protect\citeauthoryear{{Desmond} \& {Wechsler}}{{Desmond} \&
  {Wechsler}}{2017}]{Desmond_FJFP}
{Desmond} H.,  {Wechsler} R.~H.,  2017, \mn@doi [\mnras]
  {10.1093/mnras/stw2804}, \href
  {https://ui.adsabs.harvard.edu/abs/2017MNRAS.465..820D} {465, 820}

\bibitem[\protect\citeauthoryear{{Desmond}, {Ferreira}, {Lavaux}  \&
  {Jasche}}{{Desmond} et~al.}{2018}]{Desmond_reconstructing}
{Desmond} H.,  {Ferreira} P.~G.,  {Lavaux} G.,   {Jasche} J.,  2018, \mn@doi
  [\mnras] {10.1093/mnras/stx3062}, \href
  {https://ui.adsabs.harvard.edu/abs/2018MNRAS.474.3152D} {474, 3152}

\bibitem[\protect\citeauthoryear{{Desmond}, {Bartlett}  \&
  {Ferreira}}{{Desmond} et~al.}{2023}]{ESR-RAR}
{Desmond} H.,  {Bartlett} D.~J.,   {Ferreira} P.~G.,  2023, \mn@doi [\mnras]
  {10.1093/mnras/stad597}, \href
  {https://ui.adsabs.harvard.edu/abs/2023MNRAS.521.1817D} {521, 1817}

\bibitem[\protect\citeauthoryear{{Di Cintio} \& {Lelli}}{{Di Cintio} \&
  {Lelli}}{2016}]{DC_Lelli}
{Di Cintio} A.,  {Lelli} F.,  2016, \mn@doi [\mnras] {10.1093/mnrasl/slv185},
  \href {http://adsabs.harvard.edu/abs/2016MNRAS.456L.127D} {456, L127}

\bibitem[\protect\citeauthoryear{{Djorgovski} \& {Davis}}{{Djorgovski} \&
  {Davis}}{1987}]{FP_1}
{Djorgovski} S.,  {Davis} M.,  1987, \mn@doi [\apj] {10.1086/164948}, \href
  {https://ui.adsabs.harvard.edu/abs/1987ApJ...313...59D} {313, 59}

\bibitem[\protect\citeauthoryear{{Dressler}, {Lynden-Bell}, {Burstein},
  {Davies}, {Faber}, {Terlevich}  \& {Wegner}}{{Dressler} et~al.}{1987}]{FP_2}
{Dressler} A.,  {Lynden-Bell} D.,  {Burstein} D.,  {Davies} R.~L.,  {Faber}
  S.~M.,  {Terlevich} R.,   {Wegner} G.,  1987, \mn@doi [\apj]
  {10.1086/164947}, \href
  {https://ui.adsabs.harvard.edu/abs/1987ApJ...313...42D} {313, 42}

\bibitem[\protect\citeauthoryear{{Dubois} et~al.,}{{Dubois} et~al.}{2021}]{NH}
{Dubois} Y.,  et~al., 2021, \mn@doi [\aap] {10.1051/0004-6361/202039429}, \href
  {https://ui.adsabs.harvard.edu/abs/2021A&A...651A.109D} {651, A109}

\bibitem[\protect\citeauthoryear{Elson, de Blok  \& Kraan-Korteweg}{Elson
  et~al.}{2010}]{Elson_3}
Elson E.~C.,  de Blok W. J.~G.,   Kraan-Korteweg R.~C.,  2010, \mn@doi [\mnras]
  {10.1111/j.1365-2966.2010.16422.x}, 404, 2061

\bibitem[\protect\citeauthoryear{Elson, de Blok  \& Kraan-Korteweg}{Elson
  et~al.}{2011a}]{Elson_1}
Elson E.~C.,  de Blok W. J.~G.,   Kraan-Korteweg R.~C.,  2011a, \mn@doi
  [\mnras] {10.1111/j.1365-2966.2010.17672.x}, 411, 200

\bibitem[\protect\citeauthoryear{Elson, de Blok  \& Kraan-Korteweg}{Elson
  et~al.}{2011b}]{Elson_2}
Elson E.~C.,  de Blok W. J.~G.,   Kraan-Korteweg R.~C.,  2011b, \mn@doi
  [\mnras] {10.1111/j.1365-2966.2011.18701.x}, 415, 323

\bibitem[\protect\citeauthoryear{{Faber} \& {Jackson}}{{Faber} \&
  {Jackson}}{1976}]{Faber-Jackson}
{Faber} S.~M.,  {Jackson} R.~E.,  1976, \mn@doi [\apj] {10.1086/154215}, \href
  {https://ui.adsabs.harvard.edu/abs/1976ApJ...204..668F} {204, 668}

\bibitem[\protect\citeauthoryear{{Famaey} \& {Binney}}{{Famaey} \&
  {Binney}}{2005}]{simple}
{Famaey} B.,  {Binney} J.,  2005, \mn@doi [\mnras]
  {10.1111/j.1365-2966.2005.09474.x}, \href
  {https://ui.adsabs.harvard.edu/abs/2005MNRAS.363..603F} {363, 603}

\bibitem[\protect\citeauthoryear{{Famaey} \& {McGaugh}}{{Famaey} \&
  {McGaugh}}{2012}]{Famaey_McGaugh}
{Famaey} B.,  {McGaugh} S.~S.,  2012, \mn@doi [Living Reviews in Relativity]
  {10.12942/lrr-2012-10}, \href
  {http://adsabs.harvard.edu/abs/2012LRR....15...10F} {15, 10}

\bibitem[\protect\citeauthoryear{{Freundlich}, {Famaey}, {Oria}, {B{\'\i}lek},
  {M{\"u}ller}  \& {Ibata}}{{Freundlich} et~al.}{2022}]{Freundlich}
{Freundlich} J.,  {Famaey} B.,  {Oria} P.-A.,  {B{\'\i}lek} M.,  {M{\"u}ller}
  O.,   {Ibata} R.,  2022, \mn@doi [\aap] {10.1051/0004-6361/202142060}, \href
  {https://ui.adsabs.harvard.edu/abs/2022A&A...658A..26F} {658, A26}

\bibitem[\protect\citeauthoryear{Gelman \& Rubin}{Gelman \&
  Rubin}{1992}]{Gelman_Rubin}
Gelman A.,  Rubin D.~B.,  1992, \mn@doi [Statistical Science]
  {10.1214/ss/1177011136}, 7, 457

\bibitem[\protect\citeauthoryear{{Gnedin}, {Weinberg}, {Pizagno}, {Prada}  \&
  {Rix}}{{Gnedin} et~al.}{2007}]{Gnedin}
{Gnedin} O.~Y.,  {Weinberg} D.~H.,  {Pizagno} J.,  {Prada} F.,   {Rix} H.-W.,
  2007, \mn@doi [\apj] {10.1086/523256}, \href
  {https://ui.adsabs.harvard.edu/abs/2007ApJ...671.1115G} {671, 1115}

\bibitem[\protect\citeauthoryear{{Gopika} \& {Desai}}{{Gopika} \&
  {Desai}}{2021}]{groups}
{Gopika} K.,  {Desai} S.,  2021, \mn@doi [Physics of the Dark Universe]
  {10.1016/j.dark.2021.100874}, \href
  {https://ui.adsabs.harvard.edu/abs/2021PDU....3300874G} {33, 100874}

\bibitem[\protect\citeauthoryear{{Hadzhiyska}, {Wolz}, {Azzoni}, {Alonso},
  {Garc{\'\i}a-Garc{\'\i}a}, {Ruiz-Zapatero}  \& {Slosar}}{{Hadzhiyska}
  et~al.}{2023}]{David}
{Hadzhiyska} B.,  {Wolz} K.,  {Azzoni} S.,  {Alonso} D.,
  {Garc{\'\i}a-Garc{\'\i}a} C.,  {Ruiz-Zapatero} J.,   {Slosar} A.,  2023,
  \mn@doi [arXiv e-prints] {10.48550/arXiv.2301.11895}, \href
  {https://ui.adsabs.harvard.edu/abs/2023arXiv230111895H} {p. arXiv:2301.11895}

\bibitem[\protect\citeauthoryear{{Haghi} et~al.,}{{Haghi} et~al.}{2019}]{Haghi}
{Haghi} H.,  et~al., 2019, \mn@doi [\mnras] {10.1093/mnras/stz1465}, \href
  {https://ui.adsabs.harvard.edu/abs/2019MNRAS.487.2441H} {487, 2441}

\bibitem[\protect\citeauthoryear{Hernandez \& Lara-D~I}{Hernandez \&
  Lara-D~I}{2019}]{Hernandez_Lara}
Hernandez X.,  Lara-D~I A.~J.,  2019, \mn@doi [\mnras] {10.1093/mnras/stz3038},
  491, 272

\bibitem[\protect\citeauthoryear{{Hernandez}, {Cort{\'e}s}, {Allen}  \&
  {Scarpa}}{{Hernandez} et~al.}{2019}]{Hernandez}
{Hernandez} X.,  {Cort{\'e}s} R.~A.~M.,  {Allen} C.,   {Scarpa} R.,  2019,
  \mn@doi [International Journal of Modern Physics D]
  {10.1142/S0218271819501013}, \href
  {https://ui.adsabs.harvard.edu/abs/2019IJMPD..2850101H} {28, 1950101}

\bibitem[\protect\citeauthoryear{Hernandez, Cookson  \& Cortés}{Hernandez
  et~al.}{2021}]{Hernandez_Cookson_Cortes}
Hernandez X.,  Cookson S.,   Cortés R. A.~M.,  2021, \mn@doi [\mnras]
  {10.1093/mnras/stab3038}, 509, 2304

\bibitem[\protect\citeauthoryear{{Hoffman} \& {Gelman}}{{Hoffman} \&
  {Gelman}}{2011}]{NUTS}
{Hoffman} M.~D.,  {Gelman} A.,  2011, \mn@doi [arXiv e-prints]
  {10.48550/arXiv.1111.4246}, \href
  {https://ui.adsabs.harvard.edu/abs/2011arXiv1111.4246H} {p. arXiv:1111.4246}

\bibitem[\protect\citeauthoryear{{Katz}, {McGaugh}, {Sellwood}  \& {de
  Blok}}{{Katz} et~al.}{2014}]{Harley_asym}
{Katz} H.,  {McGaugh} S.~S.,  {Sellwood} J.~A.,   {de Blok} W.~J.~G.,  2014,
  \mn@doi [\mnras] {10.1093/mnras/stu070}, \href
  {https://ui.adsabs.harvard.edu/abs/2014MNRAS.439.1897K} {439, 1897}

\bibitem[\protect\citeauthoryear{{Keller} \& {Wadsley}}{{Keller} \&
  {Wadsley}}{2017}]{Keller}
{Keller} B.~W.,  {Wadsley} J.~W.,  2017, \mn@doi [\apjl]
  {10.3847/2041-8213/835/1/L17}, \href
  {https://ui.adsabs.harvard.edu/abs/2017ApJ...835L..17K} {835, L17}

\bibitem[\protect\citeauthoryear{{Lelli}, {McGaugh}  \& {Schombert}}{{Lelli}
  et~al.}{2016}]{SPARC}
{Lelli} F.,  {McGaugh} S.~S.,   {Schombert} J.~M.,  2016, \mn@doi [\aj]
  {10.3847/0004-6256/152/6/157}, \href
  {http://adsabs.harvard.edu/abs/2016AJ....152..157L} {152, 157}

\bibitem[\protect\citeauthoryear{{Lelli}, {McGaugh}, {Schombert}  \&
  {Pawlowski}}{{Lelli} et~al.}{2017}]{RAR}
{Lelli} F.,  {McGaugh} S.~S.,  {Schombert} J.~M.,   {Pawlowski} M.~S.,  2017,
  \mn@doi [\apj] {10.3847/1538-4357/836/2/152}, \href
  {https://ui.adsabs.harvard.edu/abs/2017ApJ...836..152L} {836, 152}

\bibitem[\protect\citeauthoryear{{Li}, {Lelli}, {McGaugh}  \& {Schombert}}{{Li}
  et~al.}{2018}]{Li}
{Li} P.,  {Lelli} F.,  {McGaugh} S.,   {Schombert} J.,  2018, \mn@doi [\aap]
  {10.1051/0004-6361/201732547}, \href
  {https://ui.adsabs.harvard.edu/abs/2018A&A...615A...3L} {615, A3}

\bibitem[\protect\citeauthoryear{{Ludlow} et~al.,}{{Ludlow}
  et~al.}{2017}]{Ludlow}
{Ludlow} A.~D.,  et~al., 2017, \mn@doi [\prl] {10.1103/PhysRevLett.118.161103},
  \href {https://ui.adsabs.harvard.edu/abs/2017PhRvL.118p1103L} {118, 161103}

\bibitem[\protect\citeauthoryear{{McGaugh}}{{McGaugh}}{2004}]{MDAR_McGaugh}
{McGaugh} S.~S.,  2004, \mn@doi [\apj] {10.1086/421338}, \href
  {https://ui.adsabs.harvard.edu/abs/2004ApJ...609..652M} {609, 652}

\bibitem[\protect\citeauthoryear{{McGaugh} \& {Milgrom}}{{McGaugh} \&
  {Milgrom}}{2013}]{McGaugh_Milgrom}
{McGaugh} S.,  {Milgrom} M.,  2013, \mn@doi [\apj]
  {10.1088/0004-637X/775/2/139}, \href
  {https://ui.adsabs.harvard.edu/abs/2013ApJ...775..139M} {775, 139}

\bibitem[\protect\citeauthoryear{{McGaugh} \& {Schombert}}{{McGaugh} \&
  {Schombert}}{2014}]{McGaugh-Schombert}
{McGaugh} S.~S.,  {Schombert} J.~M.,  2014, \mn@doi [\aj]
  {10.1088/0004-6256/148/5/77}, \href
  {https://ui.adsabs.harvard.edu/abs/2014AJ....148...77M} {148, 77}

\bibitem[\protect\citeauthoryear{{McGaugh} \& {Wolf}}{{McGaugh} \&
  {Wolf}}{2010}]{McGaugh_Wolf}
{McGaugh} S.~S.,  {Wolf} J.,  2010, \mn@doi [\apj]
  {10.1088/0004-637X/722/1/248}, \href
  {https://ui.adsabs.harvard.edu/abs/2010ApJ...722..248M} {722, 248}

\bibitem[\protect\citeauthoryear{{McGaugh}, {Schombert}, {Bothun}  \& {de
  Blok}}{{McGaugh} et~al.}{2000}]{McGaugh_BTFR}
{McGaugh} S.~S.,  {Schombert} J.~M.,  {Bothun} G.~D.,   {de Blok} W.~J.~G.,
  2000, \mn@doi [\apjl] {10.1086/312628}, \href
  {https://ui.adsabs.harvard.edu/abs/2000ApJ...533L..99M} {533, L99}

\bibitem[\protect\citeauthoryear{McGaugh, Lelli  \& Schombert}{McGaugh
  et~al.}{2020}]{research_note}
McGaugh S.~S.,  Lelli F.,   Schombert J.~M.,  2020, \mn@doi [Research Notes of
  the AAS] {10.3847/2515-5172/ab8471}, 4, 45

\bibitem[\protect\citeauthoryear{{Meidt} et~al.,}{{Meidt} et~al.}{2014}]{Meidt}
{Meidt} S.~E.,  et~al., 2014, \mn@doi [\apj] {10.1088/0004-637X/788/2/144},
  \href {https://ui.adsabs.harvard.edu/abs/2014ApJ...788..144M} {788, 144}

\bibitem[\protect\citeauthoryear{Meurer, Carignan, Beaulieu  \& Freeman}{Meurer
  et~al.}{1996}]{Meurer}
Meurer G.~R.,  Carignan C.,  Beaulieu S.,   Freeman K.~C.,  1996, \mn@doi [\aj]
  {10.1086/117895}, 111, 1551

\bibitem[\protect\citeauthoryear{{Milgrom}}{{Milgrom}}{1983a}]{Milgrom_1}
{Milgrom} M.,  1983a, \mn@doi [\apj] {10.1086/161130}, \href
  {http://adsabs.harvard.edu/abs/1983ApJ...270..365M} {270, 365}

\bibitem[\protect\citeauthoryear{{Milgrom}}{{Milgrom}}{1983b}]{Milgrom_3}
{Milgrom} M.,  1983b, \mn@doi [\apj] {10.1086/161131}, \href
  {http://adsabs.harvard.edu/abs/1983ApJ...270..371M} {270, 371}

\bibitem[\protect\citeauthoryear{{Milgrom}}{{Milgrom}}{1983c}]{Milgrom_2}
{Milgrom} M.,  1983c, \mn@doi [\apj] {10.1086/161132}, \href
  {http://adsabs.harvard.edu/abs/1983ApJ...270..384M} {270, 384}

\bibitem[\protect\citeauthoryear{{Milgrom}}{{Milgrom}}{2011}]{inertia_2}
{Milgrom} M.,  2011, \mn@doi [arXiv e-prints] {10.48550/arXiv.1111.1611}, \href
  {https://ui.adsabs.harvard.edu/abs/2011arXiv1111.1611M} {p. arXiv:1111.1611}

\bibitem[\protect\citeauthoryear{{Milgrom}}{{Milgrom}}{2016}]{Milgrom_2016}
{Milgrom} M.,  2016, \mn@doi [arXiv e-prints] {10.48550/arXiv.1609.06642},
  \href {https://ui.adsabs.harvard.edu/abs/2016arXiv160906642M} {p.
  arXiv:1609.06642}

\bibitem[\protect\citeauthoryear{Milgrom}{Milgrom}{2022}]{inertia_1}
Milgrom M.,  2022, \mn@doi [Phys. Rev. D] {10.1103/PhysRevD.106.064060}, 106,
  064060

\bibitem[\protect\citeauthoryear{{Navarro}, {Ben{\'\i}tez-Llambay}, {Fattahi},
  {Frenk}, {Ludlow}, {Oman}, {Schaller}  \& {Theuns}}{{Navarro}
  et~al.}{2017}]{Navarro}
{Navarro} J.~F.,  {Ben{\'\i}tez-Llambay} A.,  {Fattahi} A.,  {Frenk} C.~S.,
  {Ludlow} A.~D.,  {Oman} K.~A.,  {Schaller} M.,   {Theuns} T.,  2017, \mn@doi
  [\mnras] {10.1093/mnras/stx1705}, \href
  {https://ui.adsabs.harvard.edu/abs/2017MNRAS.471.1841N} {471, 1841}

\bibitem[\protect\citeauthoryear{{Nelson} et~al.,}{{Nelson}
  et~al.}{2019}]{TNG50_2}
{Nelson} D.,  et~al., 2019, \mn@doi [\mnras] {10.1093/mnras/stz2306}, \href
  {https://ui.adsabs.harvard.edu/abs/2019MNRAS.490.3234N} {490, 3234}

\bibitem[\protect\citeauthoryear{{Oman}, {Brouwer}, {Ludlow}  \&
  {Navarro}}{{Oman} et~al.}{2020}]{Oman}
{Oman} K.~A.,  {Brouwer} M.~M.,  {Ludlow} A.~D.,   {Navarro} J.~F.,  2020,
  arXiv e-prints, \href {https://ui.adsabs.harvard.edu/abs/2020arXiv200606700O}
  {p. arXiv:2006.06700}

\bibitem[\protect\citeauthoryear{{Paranjape} \& {Sheth}}{{Paranjape} \&
  {Sheth}}{2021}]{Paranjape_Sheth}
{Paranjape} A.,  {Sheth} R.~K.,  2021, \mn@doi [\mnras]
  {10.1093/mnras/stab2141}, \href
  {https://ui.adsabs.harvard.edu/abs/2021MNRAS.507..632P} {507, 632}

\bibitem[\protect\citeauthoryear{{Paranjape} \& {Sheth}}{{Paranjape} \&
  {Sheth}}{2022}]{PS_EFE}
{Paranjape} A.,  {Sheth} R.~K.,  2022, \mn@doi [\mnras]
  {10.1093/mnras/stac2689}, \href
  {https://ui.adsabs.harvard.edu/abs/2022MNRAS.517..130P} {517, 130}

\bibitem[\protect\citeauthoryear{Petersen \& Lelli}{Petersen \&
  Lelli}{2020}]{Peterson}
Petersen J.,  Lelli F.,  2020, \mn@doi [A\&A] {10.1051/0004-6361/201936964},
  636, A56

\bibitem[\protect\citeauthoryear{Phan, Pradhan  \& Jankowiak}{Phan
  et~al.}{2019}]{phan2019composable}
Phan D.,  Pradhan N.,   Jankowiak M.,  2019, arXiv preprint arXiv:1912.11554

\bibitem[\protect\citeauthoryear{{Pillepich} et~al.,}{{Pillepich}
  et~al.}{2019}]{TNG50_1}
{Pillepich} A.,  et~al., 2019, \mn@doi [\mnras] {10.1093/mnras/stz2338}, \href
  {https://ui.adsabs.harvard.edu/abs/2019MNRAS.490.3196P} {490, 3196}

\bibitem[\protect\citeauthoryear{{Pizagno} et~al.,}{{Pizagno}
  et~al.}{2007}]{Pizagno}
{Pizagno} J.,  et~al., 2007, \mn@doi [\aj] {10.1086/519522}, \href
  {https://ui.adsabs.harvard.edu/abs/2007AJ....134..945P} {134, 945}

\bibitem[\protect\citeauthoryear{{Pradyumna} \& {Desai}}{{Pradyumna} \&
  {Desai}}{2021}]{clusters_3}
{Pradyumna} S.,  {Desai} S.,  2021, \mn@doi [Physics of the Dark Universe]
  {10.1016/j.dark.2021.100854}, \href
  {https://ui.adsabs.harvard.edu/abs/2021PDU....3300854P} {33, 100854}

\bibitem[\protect\citeauthoryear{{Rong} et~al.,}{{Rong}
  et~al.}{2018}]{Yu_early}
{Rong} Y.,  et~al., 2018, \mn@doi [\mnras] {10.1093/mnras/sty697}, \href
  {https://ui.adsabs.harvard.edu/abs/2018MNRAS.477..230R} {477, 230}

\bibitem[\protect\citeauthoryear{{Sanders}}{{Sanders}}{1990}]{MDAR_Sanders}
{Sanders} R.~H.,  1990, \mn@doi [\aapr] {10.1007/BF00873540}, \href
  {https://ui.adsabs.harvard.edu/abs/1990A&ARv...2....1S} {2, 1}

\bibitem[\protect\citeauthoryear{{Tang} et~al.,}{{Tang} et~al.}{2022}]{Tang}
{Tang} et~al., 2022, \mn@doi [A\&A] {10.1051/0004-6361/202243944}, 668, A179

\bibitem[\protect\citeauthoryear{{Tenneti}, {Mao}, {Croft}, {Di Matteo},
  {Kosowsky}, {Zago}  \& {Zentner}}{{Tenneti} et~al.}{2018}]{Tenneti}
{Tenneti} A.,  {Mao} Y.-Y.,  {Croft} R. A.~C.,  {Di Matteo} T.,  {Kosowsky} A.,
   {Zago} F.,   {Zentner} A.~R.,  2018, \mn@doi [\mnras]
  {10.1093/mnras/stx3010}, \href
  {https://ui.adsabs.harvard.edu/abs/2018MNRAS.474.3125T} {474, 3125}

\bibitem[\protect\citeauthoryear{{Tian}, {Umetsu}, {Ko}, {Donahue}  \&
  {Chiu}}{{Tian} et~al.}{2020}]{clusters_2}
{Tian} Y.,  {Umetsu} K.,  {Ko} C.-M.,  {Donahue} M.,   {Chiu} I.~N.,  2020,
  \mn@doi [\apj] {10.3847/1538-4357/ab8e3d}, \href
  {https://ui.adsabs.harvard.edu/abs/2020ApJ...896...70T} {896, 70}

\bibitem[\protect\citeauthoryear{{Tully} \& {Fisher}}{{Tully} \&
  {Fisher}}{1977}]{Tully-Fisher}
{Tully} R.~B.,  {Fisher} J.~R.,  1977, \aap, \href
  {https://ui.adsabs.harvard.edu/abs/1977A&A....54..661T} {54, 661}

\bibitem[\protect\citeauthoryear{{Tully} et~al.,}{{Tully} et~al.}{2023}]{CF4}
{Tully} R.~B.,  et~al., 2023, \mn@doi [\apj] {10.3847/1538-4357/ac94d8}, \href
  {https://ui.adsabs.harvard.edu/abs/2023ApJ...944...94T} {944, 94}

\bibitem[\protect\citeauthoryear{{Zonoozi}, {Lieberz}, {Banik}, {Haghi}  \&
  {Kroupa}}{{Zonoozi} et~al.}{2021}]{EFE_QUMOND}
{Zonoozi} A.~H.,  {Lieberz} P.,  {Banik} I.,  {Haghi} H.,   {Kroupa} P.,  2021,
  \mn@doi [\mnras] {10.1093/mnras/stab2068}, \href
  {https://ui.adsabs.harvard.edu/abs/2021MNRAS.506.5468Z} {506, 5468}

\makeatother
\end{thebibliography}


\label{lastpage}
\end{document}